\documentclass{emulateapj}

\usepackage{color}
\usepackage{amsmath}
\usepackage{graphicx}
\usepackage{threeparttable}

\newcommand{\mr}{\mathrm} 

\newcommand{\sgmg}{\Sigma_\mr{gas}} 
\newcommand{\rmol}{R_\mr{mol}} 
\newcommand{\sgms}{\Sigma_\mr{star}} 
\newcommand{\sgmsf}{\Sigma_\mr{SFR}} 
\newcommand{\sgmsfnet}{\Sigma_\mr{SFR,net}} 
\newcommand{\sgmin}{\Sigma_\mr{in}} 
\newcommand{\sgmout}{\Sigma_\mr{out}} 
\newcommand{\sgmre}{\Sigma_\mr{re}} 
\newcommand{\sgmrez}{\Sigma_\mr{re,0}} 
\newcommand{\ret}{\mathcal{R}} 
\newcommand{\tr}{(t,\:R)} 
\newcommand{\rf}{R_\mr{f}} 
\newcommand{\srm}{\sigma_\mr{RM}} 

\newcommand{\mtotin}{M_\mr{tot,in}} 
\newcommand{\hrino}{h_\mr{R,in1}} 
\newcommand{\hrint}{h_\mr{R,in2}} 
\newcommand{\rb}{R_\mr{b}} 
\newcommand{\tauin}{\tau_\mr{in}} 
\newcommand{\tauinz}{\tau_\mr{in,0}} 
\newcommand{\tauine}{\tau_\mr{in,8}} 

\newcommand{\lmdz}{\Lambda_0} 
\newcommand{\lmde}{\Lambda_8} 

\newcommand{\fre}{f_\mr{re}} 
\newcommand{\tre}{t_\mr{re}} 
\newcommand{\kre}{k_\mr{re}} 

\newcommand{\arm}{a_\mr{rm}} 
\newcommand{\brm}{b_\mr{rm}} 
\newcommand{\crm}{c_\mr{rm}} 
\newcommand{\drm}{d_\mr{rm}} 

\newcommand{\fia}{f_\mr{Ia}} 

\newcommand{\afe}{{\rm [\alpha/Fe]}} 
\newcommand{\feh}{{\rm [Fe/H]}} 

\newcommand{\msun}{{\rm M_\odot}} 

\newcommand{\krebest}{0.25} 
\newcommand{\armbest}{0.13} 
\newcommand{\brmbest}{1.7} 
\newcommand{\gnmbest}{0.24} 

\accepted{February, 15, 2018}

\shorttitle{Metallicity distribution of disk stars and the formation history of the Milky Way}
\shortauthors{Toyouchi \& Chiba}

\begin{document}

\title{Metallicity Distribution of Disk Stars and the Formation History of the Milky Way}

\author{Daisuke~Toyouchi\altaffilmark{1,2} and
	Masashi~Chiba\altaffilmark{2}}

\altaffiltext{1}{Theoretical Astrophysics Group, Department of Physics, Kyoto University,
Sakyo-ku, Kyoto 606-8502, Japan}
\altaffiltext{2}{Astronomical Institute, Tohoku University,
Aoba-ku, Sendai 980-8578, Japan}


\begin{abstract}

We investigate the formation history of the stellar disk component in the Milky Way (MW) based on our new chemical evolution model. Our model considers several fundamental baryonic processes, including gas infall, re-accretion of outflowing gas, and radial migration of disk stars. Each of these baryonic processes in the disk evolution is characterized by model parameters, which are determined by fitting to various observational data of the stellar disk in the MW, including the radial dependence of the metallicity distribution function (MDF) of the disk stars, which has recently been derived in the APOGEE survey. We succeeded to obtain the best set of model parameters, which well reproduces the observed radial dependences of the mean, standard deviation, skewness, and kurtosis of the MDFs for the disk stars. We analyze the basic properties of our model results in detail to get new insights into the important baryonic processes in the formation history of the MW. One of the remarkable findings is that outflowing gas, containing much heavy elements, preferentially re-accretes onto the outer disk parts, and this recycling process of metal-enriched gas is a key ingredient to reproduce the observed narrower MDFs at larger radii. Moreover, important implications for the radial dependence of gas infall and the influence of radial migration on the MDFs are also inferred from our model calculation. Thus, the MDF of disk stars is a useful clue for studying the formation history of the MW.

\end{abstract}


\section{Introduction} \label{sec:intro}

Disk galaxies are a dominant galaxy system in the present universe (Delgado-Serrano et al. 2010), so that unraveling their formation and evolution histories is a major subject in galactic astronomy. Formation of disk galaxies complexly involves various baryonic processes. In particular, gas infall onto the galactic disk plane is an essential process, which dominates the budget of baryon and star formation activity in disk galaxies. In the grand picture of galaxy formation and evolution based on $\Lambda$CDM cosmology, gas infall onto a galactic disk occurs when gas accretes from the circumgalactic region into a dark matter halo and subsequently cools and collapses toward its central region. The gas cooling and collapsing processes in disk galaxies are significantly affected by various feedback processes associated with, e.g., UV radiation from massive stars, supernovae explosions, and active galactic nuclei. 

Such feedback processes are also important as a driver of a strong galactic gas outflow. Several works suggest that some parts of outflowing gas should escape and be eternally lost from star-forming galaxies to explain their observed chemical properties (e.g., Peeples et al. 2014; Yabe et al. 2015; Toyouchi \& Chiba 2015). However, most of the outflowing gas are expected to fall back onto the galactic disk plane (e.g., Oppenheimer et al. 2010): according to the numerical simulation by Oppenheimer \& Dav\'e (2008), the fraction of re-accreting gas to the total outflowing gas is over 80 \% in MW-sized halos at z $\sim$ 0. Moreover, the other simulations suggest that metal-enriched gas ejected from inner disk regions acquires additional angular momentum via feedback processes and preferentially re-accretes onto disk regions further out than the radii where it originally resided (Bekki et al. 2009; Gibson et al. 2013). Such transportations of metal-enriched gas from inner to outer disk regions can significantly change the chemical abundance of the interstellar medium (ISM) at the outskirts of disk galaxies (Tsujimoto et al. 2010). 

Thus, gas infall, gas outflow and subsequent re-accretion of outflowing gas are very important processes, which can significantly influence the evolution of galaxies. However, predicting the actual effects of the microscopic gas cooling and heating mechanisms, which control these gas flow processes, is generally difficult in the context of galaxy evolution. Therefore, to get meaningful insights into the galactic gas infall, outflow and re-accretion processes, many theoretical efforts based on numerical simulations (e.g., Dav\'e et al. 2011; Hopkins et al. 2012; Gibson et al. 2013; Okamoto et al. 2014; Muratov et al. 2015) and semi-analytic models (e.g., Lilly et al. 2013; Zahid et al. 2014; Lu et al. 2015; Yabe et al. 2015; Toyouchi \& Chiba 2015; Zhu et al. 2016), are actively being made.

In addition to the effect of gas flow processes, radial migration of stars along a galactic disk affects the structure and dynamics of the disk galaxy significantly. In fact, various theoretical and observational studies have suggested the importance of radial redistributions of disk stars (e.g., Sch\"onrich \& Binney 2009; Hayden et al. 2015; Kordopatis et al. 2015; Morishita et al. 2015). Such radial migration processes are generally triggered by gravitational interactions between disk stars and bar/spiral structures and giant molecular clouds (GMCs) (e.g., Sellwood \& Binney 2002; Ro\'skar et al. 2008; Loebman et al. 2011), and minor mergers of satellite galaxies (e.g., Quinn et al. 1993; Vel\'azquez \& White 1999; Villalobos \& Helmi 2008). The basic properties of radial migration depend significantly on its triggering process. Therefore, the detailed investigation of radial migration history in a disk galaxy provides important implications for the internal and external gravitational perturbations working in the galaxy.

For studying these important baryonic processes in disk galaxy evolution, the MW is the best site, because we can observe individual stars composing the stellar disk in great detail. Many observations of the MW disk stars have shown the various detailed properties of the present stellar disk, such as the spatial structure (e.g., Yoshii 1982; Gilmore \& Reid 1983; Juri\'c et al. 2008; Bovy et al. 2012, 2016), [$\alpha$/Fe]$-$metallicity relation (e.g., Bensby et al. 2003, 2014; Lee et al. 2011; Adibekyan et al. 2012), and radial metallicity gradient (e.g., Nordstr\"om et al. 2004; Allende Prieto et al. 2006; Cheng et al. 2012; Toyouchi \& Chiba 2014). These observed properties of the Galactic stellar disk are useful constraints on its formation history. 

In particular the metallicity distribution function (MDF) of the Galactic disk stars is a very tight constraint for studying the formation history of the MW, because its properties are very sensitive to the past star formation, chemical evolution, and radial migration histories in the Galactic disk (e.g., Pagel 1989; Chiappini et al. 1997, 2001; Sch\"onrich \& Binney 2009; Loebman et al. 2016). The MDF of disk stars has been actively investigated in various spectroscopic observations since 1990s (e.g., Wyse \& Gilmore 1995; Lee et al. 2011). Recently, the observation with SDSS-III/APOGEE, which is one of the latest large surveys for the MW stellar disk, revealed the MDFs in the unprecedented wide radial range, $R = 3-16$ kpc, as shown in Figure \ref{fig:mdf} (Anders et al. 2014; Hayden et al. 2015). 

Investigating the origin of the observed radial dependence of the MDFs with semi-analytic models for studying chemical evolution of disk galaxies will provide new important implications for the formation histories of the Galactic stellar disk. Such chemical evolution models have been studied in many previous papers (e.g., Sch\"onrich \& Binney 2009; Kubryk et al. 2015a; Toyouchi \& Chiba 2016). In this work, we present a new semi-analytic chemical evolution model, which explicitly takes into account gas infall, outflow, re-accretion, and radial migration. Then, by combining it with the parameter surveys based on the Markov Chain Monte Carlo (MCMC) method, we attempt to derive the best solution for galaxy formation and evolution, which reproduces the observed radial dependence of the MDF of the Galactic disk stars. Finally, based on our model experiments, we discuss the gas infall, gas outflow, re-accretion of outflowing gas, and radial migration histories of the MW.

We should note here that recently Loebman et al. (2016) discussed the MDFs of disk stars observed in the APOGEE survey with the isolated hydrodynamical numerical simulation, and showed that the radial change of the skewness of the MDFs can be naturally explained as a result of stellar migration without any fine tunings. However, they did not present the conditions of gas infall and outflow to produce the observed radial dependence of the MDFs. Therefore, our model calculation based on a semi-analytical method will be complementary to their numerical simulation. Additionally, as will be introduced in Section \ref{sec:model}, our model ensures to reproduce not only the MDFs of disk stars, but also the radial distributions of disk stars, gas, and chemical abundances of the interstellar medium (ISM), observed in the Galactic disk. Thus, our semi-analytical model will be useful for studying the evolution history of the MW.

This paper is organized as follows. In Section \ref{sec:model}, we introduce our semi-analytic chemical evolution model. In Section \ref{sec:result}, we show the fitting results to the various properties of the present Galactic stellar disk based on the MCMC method. In Section \ref{sec:discussion}, we discuss our model calculation results in detail and present new implications for the formation history of the MW. In Section \ref{sec:other}, we show the validity of our model calculation. Finally, our conclusions are drawn in Section \ref{sec:summary}.

\section{Chemical Evolution Model} \label{sec:model}

In this paper, we adopt a standard one-dimensional chemical evolution model along a galactic disk, as studied in our previous work (Toyouchi \& Chiba 2016). In this model, we calculate baryonic mass evolution for a radial range from $R$ = 0 to $R_\mr{out}$ (= 16 kpc) with a grid of $\Delta R$ = 1 kpc over $t$ = 0 to $t_\mr{p}$ (= 12 Gyr) with a grid of $\Delta t$ = 50 Myr. This model calculation can provide the surface density of gas, $\sgmg$, that of stars, $\sgms$, and a mass fraction of heavy element $i$, $Z_i$, at any time, $t$, and at any radius, $R$. Additionally, we invetigate the $\feh$ and $\afe$ distributions of disk stars and their evolution, where $\afe$ is defined as the average for the typical $\alpha$ elements of O, Mg, Si, Ca, and Ti relative to the Fe abundance.

In Section 2.1-2.4, we introduce each important process in our chemical evolution model. Basic equations to calculate baryonic mass evolution in a galactic disk are described in Section 2.5. Finally, in Section 2.6 we discuss how we select the values of the important free parameters in our model. 

\subsection{Gas Infall Rate}

Infall of gas from the circumgalactic region into a dark matter halo and finally into a disk plane is an essential process in the course of galaxy formation. In our model, we assume that the surface density of gas infall rate, $\sgmin$, as functions of $t$ and $R$ is given as, 
\begin{eqnarray}
\sgmin \tr = \scalebox{1.0}{$\displaystyle \frac{\Sigma_\mr{in,0} \: t}{\tauin^2 \left \{ 1- (1+t_\mr{p}/\tauin) \mr{exp} (-t_\mr{p}/\tauin) \right \}}$} \nonumber \\
\times \begin{cases}
\scalebox{1.0}{$\displaystyle \mr{exp} \left ( -\frac{R}{\hrino} - \frac{t}{\tauin} \right )$} & (R \leq \rb) \\[11pt]
\scalebox{1.0}{$\displaystyle \mr{exp} \left ( -\frac{R-\rb}{\hrint} -\frac{\rb}{\hrino} - \frac{t}{\tauin} \right )$} & (R > \rb) \ ,
\end{cases}
\label{eq:sgmin}
\end{eqnarray} 
where $\Sigma_\mr{in,0}$ is determined such that the total mass of infalling gas on the disk plane within $R_\mr{out}$ by $t = t_\mr{p}$ equals to $\mtotin$.  In this model, we assume $\mtotin = 4.5 \times 10^{10} \ \mr{M_\odot}$, roughly corresponding to the sum of the observed stellar and gas mass of the Galactic disk at the current time, and this assumption implies that our model does not allow a significant mass loss via galactic outflow process as noted in detail in Section 2.3.  $\rb$ is the break radius where the scale length of the radial profile of gas infall changes from $\hrino$ at inner radii to $\hrint$ at outer radii. In our previous model shown in Toyouchi \& Chiba (2016), we adopted a single exponential profile of gas infall without any break. However, van den Bosch et al. (2001) based on a theoretical consideration suggest that the radial profile of gas infall can be more complex, as detailed later in Section 4.2, and therefore in this work we consider $\rb$, $\hrino$, and $\hrint$ to take into account a broken radial profile of gas infall.

$\tauin$ is a time scale of gas infall rate at $R$, described with $\tauinz$, $\tauine$ and $\alpha$ in our model as,
\begin{eqnarray}
\tauin (R) = \tauinz + (\tauine-\tauinz) \left ( \frac{R}{8 \ \mr{kpc}} \right )^{\alpha} \ .
\end{eqnarray}
Such an assumption of a radially dependent time scale of gas infall has been adopted in many previous studies (e.g., Chiappini et al. 1997, 2001) and is motivated from the radial dependence of the collapsing time scale of gas in the host halo. Additionally, it is worth noting here that $\sgmin \propto t \; \mr{exp}(-t/\tauin)$ in equation (\ref{eq:sgmin}) supposes that at a fixed radius $\sgmin$ linearly increases with increasing time up to at $t = \tauin$, and after this time exponentially decreases with timescale of $\tauin$. Such a time dependence of $\sgmin$ may be somewhat more realistic than that of $\sgmin \propto \mr{exp}(-t/\tauin)$ as usually assumed in many previous works (e.g., Chiappini et al. 1997), where at all radii $\sgmin$ is maximum at $t = 0$. This is because the time taking for newly accreting gas, possibly in the form of cold accretion from outside the halo, to reach disk regions may be finite, depending on the collapsing time of gas in the halo (van den Bosch 2001; 2002). In this work, we have carried out model calculations adopting not only $\sgmin$ in equation (\ref{eq:sgmin}), but also a different form of $\sgmin \propto \mr{exp}(-t/\tauin)$, and found that both cases provide basically the same conclusion, although in the later case, $\tauine$ is needed to be much longer than the Hubble time to reproduce the observed MDF, which significantly disagrees with the previous model predictions (Moll{\'a} \& D{\'{\i}}az 2005; Moll{\'a} et al. 2016).

In summary, the gas infall history in our chemical evolution model is characterized by the six free parameters ($\rb$, $\hrino$, $\hrint$, $\tauinz$, $\tauine$, $\alpha$).

\subsection{Star Formation Rate}

In our model, to calculate the surface density of star formation rate, $\sgmsf$, we adopt the star formation law confirmed in the observation of the local star-forming galaxies by Bigiel et al. (2008), in which $\sgmsf$ is proportional to the surface density of H$_2$ gas, as follows,
\begin{eqnarray}
\sgmsf = 1.6 \  \frac{\rmol}{\rmol+1} \  \left ( \frac{\sgmg}{M_\odot \mr{pc}^{-2}} \right )  \ [M_\odot \mr{pc}^{-2} \mr{Gyr}^{-1}] \ ,
\label{eq:sfr}
\end{eqnarray}
where $\rmol$ is the mass ratio of H$_2$ to HI gas, which can be derived by using the following semi-empirical law of $\rmol$ provided by Blitz \& Rosolowsky (2006),
\begin{eqnarray}
\rmol = 0.23 \ \left [ \left ( \frac{\sgmg}{10 \ M_\odot \mr{pc}^{-2}} \right ) \left ( \frac{\sgms}{35 \ M_\odot \mr{pc}^{-2}} \right )^{0.5} \right ]^{0.92} \ .
\label{eq:rmol}
\end{eqnarray}
These empirical laws of equations (\ref{eq:sfr}) and (\ref{eq:rmol}) have been known to reproduce the present spatial distribution of HI, H$_2$ and star formation in the MW (Blitz \& Rosolowsky 2006; Kubryk et al. 2015a).

\subsection{Gas Outflow and Re-accretion processes}

Feedback processes associated with star formation activity, such as radiation pressure from massive stars and supernova explosions, may drive a galactic gas outflow. In our model, the surface density of gas outflow rate, $\sgmout$, is assumed to be proportional to $\sgmsf$ with a proportional coefficient, $\Lambda$, which is generally called an outflow-mass loading factor. In this model we consider $\Lambda$ as a function of $R$, described as follows,
\begin{eqnarray}
\Lambda \tr = \left \{ \lmdz + (\lmde-\lmdz) \left ( \frac{R}{8 \ \mr{kpc}} \right )^{\beta} \right \}  \ ,
\label{eq:outflow}
\end{eqnarray}
where $\lmdz$, $\lmde$, $\beta$ are parameters characterizing the radial dependence of $\Lambda$. 

A fraction of gas ejected from the disk plane as outflow cannot escape from the host halo and eventually re-accretes onto the galactic disk plane. We attempt to explicitly take into account such a re-accretion process in our model. Our model supposes that the mass fraction, $\fre$, of the total outflowing gas, which ejected from the entire galactic disk at any time, falls back onto any place of the disk after the time, $\tre$. Moreover, we assume that the radial profile of surface re-accretion rate density, $\sgmre \tr$, is more radially extended than gas outflow rate at $t-\tre$, $\sgmout (t-\tre, \ R)$. This assumption is motivated by the numerical simulation by Bekki et al. (2009), showing that because gas ejected from the disk gains additional angular momentum via feedback processes, this ejected gas preferentially re-accretes into disk regions further out than the radii where it originally resided. Therefore, in our model, we give $\sgmre$ as follows,
\begin{eqnarray}
\sgmre \tr = \sgmrez (t) \ \left ( \frac{\sgmout(t-\tre, \ R)}{\sgmout(t-\tre, \ 0)} \right )^{\kre} ,
\label{eq:reacc}
\end{eqnarray}
\begin{eqnarray}
\sgmrez (t) = \nonumber \ \ \ \ \ \ \ \ \ \ \ \ \ \ \ \ \ \ \ \ \ \ \ \  \ \ \ \ \ \ \ \ \ \ \  \ \ \ \ \ \ \ \ \ \ \  \ \ \ \ \ \ \ \ \ \ \ \\ 
\frac{\fre  \int^{R_\mr{out}}_{0} R \sgmout (t-\tre, \ R) \mr{d}R}{ \int^{R_\mr{out}}_{0} R \left( \sgmout (t-\tre, \ R)/ \sgmout (t-\tre, \ 0) \right )^{\kre} \mr{d}R } \ .
\label{eq:reacc0}
\end{eqnarray}
where $\kre$ is a parameter determining the difference of the radial profile between $\sgmout$ and $\sgmre$, and here we set $0 < \kre < 1$, meaning a radial expansion of $\sgmre$ as compared to $\sgmout$. Thus, the properties of re-accretion of outflowing gas is characterized with $\fre$, $\tre$ and $\kre$ in this model.  Here, for simplicity, we deal with only $\kre$ as a free-parameter in our MCMC procedure, while we adopt fixed values for $\fre$ and $\tre$. We assume $\fre = 1$, implying that all outflowing gas fall back to the galactic disk and there is no mass loss from the host halo. This assumption is based on the result of numerical simulation by Oppenheimer \& Dave (2008), which showed that more than 80 \% of outflowing gas re-accretes onto the galactic disk and is used for the subsequent star formation. Moreover, no mass loss from the halo was assumed in the previous chemical evolution models for the MW, which can reasonably reproduce the observed mass distributions of star, gas, and heavy elements in the Galactic disk (e.g., Chiappini et al. 1997; Kubryk et al. 2015a). We also fix $\tre$ = 300 Myr, which generally corresponds to the typical time scale of falling back of gas ejected from the disk plane (e.g., Spitoni et al. 2009; Bekki et al. 2009). However, the time scale of gas re-accretion is actually unclear. In fact, Oppenheimer \& Dave (2008) predict the gas re-accretion time scale $\lesssim$ 1 Gyr for MW-like galaxies at $z \sim$ 0. Therefore, we checked how the model calculation results depend on the choice of the value of $\tre$, and found that unless $\tre$ is much longer than 1 Gyr, roughly corresponding to the dynamical time scale in the MW-like halo, our conclusion in this paper remains unchanged. 

Furthermore, we assume that the abundance of heavy element $i$ of re-accreting gas, $Z_\mr{i,re}$, at $t$ is described by using the averaged chemical abundance of outflowing gas at $t-\tre$, as follows,
\begin{eqnarray}
Z_{i,\mr{re}} (t) = \frac{\int^{R_\mr{out}}_{0} R Z_i (t-\tre, \ R) \sgmout (t-\tre, \ R) \mr{d}R}{\int^{R_\mr{out}}_{0} R \sgmout (t-\tre, \ R) \mr{d}R} \ .
\label{eq:zreacc}
\end{eqnarray}

We note here that more strictly $Z_{i,\mr{re}}$ should also depend on $R$ because the gas ejected from the inner region is expected to contain more heavy elements than those ejected from the outer one. Moreover, $\fre$ should be written as a function of $t$. Actually, Oppenheimer et al. (2010) based on the numerical simulation of galaxy formation shows that the re-accretion of outflowing gas becomes more active with the mass growth of a host halo. In this model, we adopt such a time and radially independent assumption of $\tre$ and $Z_{i,\mr{re}}$, respectively, as a first step to check the effect of re-accretion of metal-enriched outflowing gas on the chemical evolution of disk galaxies.

\subsection{Radial Migration}

In our model the effect of radial migration of disk stars is described with the method originally adopted in Sellwood \& Binney (2002), which can reproduce the basic properties of radial migration obtained in N-body simulations well. This method is based on the calculation of the probability, in which a star born at radius $\rf$ and with age $\tau$ is found in radius $R$, $P(\tau, \ \rf, \ R)$, and this is expressed in terms of the following Gaussian function 
\begin{eqnarray}
P(\tau, \ \rf, \ R) =  \frac{1}{\sqrt{2 \pi \srm^2}} \ \mr{exp} \left [ \ -\frac{(R-\rf)^2}{2 \srm^2} \ \right ] \ ,
\label{eq:p_rm}
\end{eqnarray}
where $\srm$ corresponds to the diffusion length of stars by radial migration, which is generally a function of $\tau$ and $\rf$. Thus, in this method the radial migration history can be characterized by the time dependence of $\srm$. 

In our model, we assume that $\srm$ monotonically increases with increasing $\tau$ as follows, 
\begin{eqnarray}
\srm(\tau, \ \rf) &=& (\arm \rf + \brm) \left ( \frac{\tau}{5 \ \mr{Gyr}} \right )^{\gamma} \nonumber \\
& & + (\crm \rf + \drm) \ \ [ \mr{kpc} ] \ .
\label{eq:crm}
\end{eqnarray}
where, $\arm$, $\brm$, $\crm$, $\drm$, and $\gamma$ are parameters characterizing the radial migration history. This description of $\srm$ is the same as that adopted in Kubryk et al. (2013), who analyzed the radial migration history in a bar-dominated disk galaxy in the high-resolution N-body+smoothed particle hydrodynamics simulation, and determined ($\arm$, $\brm$, $\crm$, $\drm$, $\gamma$) = ($-$0.0667, 2.75 kpc, $-$0.226, 2.71, 0.5). We note here that their choice of the values of these parameters was based on the numerical experiment for only one example of a disk galaxy. Therefore, it is worth searching for another set of these parameters to represent the radial migration history of the Galactic stellar disk based on the analysis of the observed MDFs of the disk stars. In this paper we present only the results of model calculation fixing $\crm = \drm = 0$, because a significant radial migration of zero-age stars is somewhat unphysical. Actually, even if we include $\crm$ and $\drm$ as free parameters in the MCMC procedure, the derived properties of $\srm$ are not modified significantly, although their values are somewhat degenerate with $\arm$, $\brm$, and $\gamma$. For a similar reason, for the value of $\gamma$, we adopt the lower limit of $\gamma = 0.15$ in the MCMC estimation, because without no lower limit for $\gamma$, a local minimum emerges at around $\gamma \sim 0$,  also leading to such a significant radial migration of zero-age stars.

It is also worth noting here that such a continuous evolution of $\srm$ assumed in equation (\ref{eq:crm}) supposes the radial migration events driven by internal gravitational processes, such as interactions between bar/spiral structures and disk stars. However, our recent work in Toyouchi \& Chiba (2016) suggests that not only such a continuous radial migration event, but also discontinuous one driven by an external disk heating event, such as minor merging of satellite galaxies, is important in the chemical evolution in the galactic disk. Therefore while in this study we focus on the effect of a continuous radial migration event for simplicity, more extensive modeling, including a discontinuous event, would be worth exploring as a future work.

\subsection{Basic Equations}

The basic equations in our chemical evolution model are described as follows,
\begin{eqnarray}
\frac{\partial  \sgmg}{\partial t} = - \sgmsfnet + \sgmin - \sgmout + \sgmre \ ,
\label{eq:gas}
\end{eqnarray}
\begin{eqnarray}
\sgms \tr &=& \int^{R_\mr{out}}_{0} \int^{t}_{0}  \frac{\rf}{R} \ \left (1-\ret(\tau) \right ) \nonumber \\
& & \times \sgmsf (t-\tau, \ \rf)  P(\tau, \ R, \ \rf) \mr{d}\tau  \mr{d}\rf \ , \nonumber \\
& &
\label{eq:star}
\end{eqnarray}
\begin{eqnarray}
\frac{\partial  (Z_i \sgmg)}{\partial t} &=& - P_\mr{net,i} + (Y_{\mr{II},i} + Y_{\mr{Ia},i}) \sgmsf \nonumber \\
& & + Z_{\mr{in},i} \sgmin - Z_{\mr{out},i} \sgmout + Z_{\mr{re},i} \sgmre \;.
\label{eq:metal}
\end{eqnarray}
By solving equation (\ref{eq:gas})-(\ref{eq:metal}) for each radial bin at each time step, we obtain the values of $\sgmg$, $\sgms$, and $Z_i$ at any $t$ and $R$. 

In equation (\ref{eq:gas}) the first term on the right hand side describes the net gas consumption by star formation, taking into account the effect of radial migration of stars and the delayed mass return from long-lived stars, given from the following calculation,
\begin{eqnarray}
\sgmsfnet \tr &=& \sgmsf \tr \nonumber \\
& & +  \int^{R_\mr{out}}_{0} \int^{t}_{0} \frac{\rf}{R}  \left (m_\tau - M_\mr{R} (m_\tau) \right )  \nonumber \\
& & \times  \sgmsf (t-\tau, \ \rf) \Phi(m_\tau)  P(\tau, \ R, \ \rf) \nonumber \\
& & \times \frac{\mr{d}m_\tau}{\mr{d}\tau}\mr{d}\tau  \mr{d}\rf  \ , \nonumber \\
& &
\label{eq:sfrnet}
\end{eqnarray}
where, $m_\tau$, $M_\mr{R} (m)$, and $\Phi (m)$ are the mass of stars, whose lifetime is $\tau$, the remnant mass of stars with initial mass of $m$, and initial mass function (IMF) of stars, respectively. We note that although the second terms in the right hand side of equation (\ref{eq:sfrnet}) has the positive sign, this actually can express the stellar mass loss because $\mr{d}m_\tau/\mr{d}\tau$ is always negative. To obtain $m_\tau$ and $M_\mr{R} (m)$, we adopt the mass-lifetime relation of Schaller et al. (1992) and the initial mass-remnant mass relation of Kalirai et al. (2008). In this paper, we adopt the IMF of Kroupa et al. (1993) for our model calculation. It is worth noting here that galaxy formation depends strongly on the assumed IMF, because a different IMF provides much different nucleosynthetic yields. For instance, the IMFs of Salpeter (1955) and Chabrier (2003) provide two and three times lager stellar yields of Type II SNe, respectively, than those of Kroupa et al. (1993), leading to an over-prediction of heavy elements in the Solar neighborhood (Vincenzo et al. 2016). Consequently, for these IMFs, we must adopt the assumption of $\fre < 1$, which corresponds to non-zero mass loss via galactic outflow process. Investigating such a dependence of the results on the assumed IMF is out of our interests in this paper, and therefore we here focus on the results of model calculation based on the IMF of Kroupa et al. (1993) and keep the assumption of $\fre$ = 1.

Equation (\ref{eq:star}) represents the time evolution of $\sgms$ at any radius, reflecting the past star formation and radial migration history over the galactic disk. Here, $\ret (\tau)$ means a fraction of mass, which has been returned into the ISM from stars with age of $\tau$, defined as follows, 
\begin{eqnarray}
\ret (\tau) = \int^{100 \msun}_{m_\tau} \left (m - M_\mr{R} (m) \right ) \Phi (m)  \mr{d} m \ .
\label{eq:ret}
\end{eqnarray}

In equation (\ref{eq:metal}), the first term on the right hand side describes the net loss of mass of heavy elements from the ISM due to being locked up into stars, given as follows,
\begin{eqnarray}
P_\mr{net} \tr &=& Z_i \tr \sgmsf \tr \nonumber \\
& & +  \int^{R_\mr{out}}_{0} \int^{t}_{0} \frac{\rf}{R} \ \left (m_\tau - M_\mr{R} \left(m_\tau \right) \right ) \nonumber \\
& & \times Z_i (t-\tau, \ \rf) \sgmsf (t-\tau, \ \rf) \nonumber \\
& & \times \Phi(m_\tau) P(\tau, \ R, \ \rf)  \frac{\mr{d}m_\tau}{\mr{d}\tau} \mr{d}\tau  \mr{d}\rf \ .
\label{eq:pnet}
\end{eqnarray}

The second term on the right hand side of equation (\ref{eq:metal}) is the supply of heavy element $i$ newly synthesized in stars, where $Y_{\mr{II},i}$ and $Y_{\mr{Ia},i}$ are the nucleosynthetic yields from Type II and Ia SNe (hereafter SN II and SN Ia), respectively. Here, $Y_{\mr{II},i}$ is a constant value, derived from the SN II yield for each element provided by Fran\c{c}ois et al. (2004). On the other hand, $Y_{\mr{Ia},i}$ can change with time and radius, depending on the past star formation and radial migration history as follows, 
\begin{eqnarray}
Y_{\mr{Ia},i} &=& \frac{m_{\mr{Ia},i} \ \fia}{\sgmsf \tr} \int^{R_\mr{out}}_{0} \int^{t}_{\Delta t_\mr{Ia}}  \frac{\rf}{R} \nonumber \\
& & \times \frac{\sgmsf (t-\tau, \ \rf)  P(\tau, \ R, \ \rf)}{\tau}  \mr{d}\tau  \mr{d}\rf \ ,
\label{eq:yia}
\end{eqnarray}
where $\fia$ is a free parameter controlling the SN Ia rate in the galactic disk, and $m_{\mr{Ia},i}$ is the released mass of heavy element $i$ per a Type Ia supernova, for which we adopt the SN Ia yield of W7 model in Iwamoto et al. (1999). $\Delta t_\mr{Ia}$ is a minimum delayed time of SN Ia. Here we set $\Delta t_\mr{Ia}$ = 0.5 Gyr, which is suggested by Homma et al. (2015) to reproduce star formation histories and chemical evolutions of the Galactic dwarf galaxies self-consistently. It is worth noting here that while the value of $\Delta t_\mr{Ia}$ in our model is different from the general one adopted in most of previous studies of $\Delta t_\mr{Ia}$ = 0.1 Gyr, the main results and conclusions in this paper are independent of the choice of the value of $\Delta t_\mr{Ia}$.

The third and fourth terms on the right hand side of equation (\ref{eq:metal}) denote the mass injection and ejection of heavy elements associated with infall and outflow, respectively, where $Z_{\mr{in},i}$ and $Z_{\mr{out},i}$ are mass fractions of heavy element $i$ in infalling and outflowing gas, respectively. In this model, we set the iron abundance of infalling gas of $\feh = -1$, which is in good agreement with the low-metal edge of the distribution function of $\feh$ for the whole disk stars observed in APOGEE survey (Hayden et al. 2015), and therefore in our model calculation any different choice of the abundance of infalling gas fails to reproduce the observed MDFs. In addition, for the abundances of $\alpha$ elements of infalling gas we set $\afe = 0.4$, which roughly equals to the average abundance ratio of $Y_\mr{II,\alpha}$ to $Y_\mr{II, Fe}$.  We assume that the metallicity of outflowing gas always equals to that of the ISM, implying $Z_{\mr{out},i} = Z_i$. This assumption is generally reasonable because the outflowing gas is expected to mainly consists of the ISM, which acquired much momentum or kinetic energy via radiations from massive stars and SN explosions, rather than the direct ejecta of SNe including much heavy elements.
 
\subsection{Determination of Model Parameters}

\begin{table}
\caption{The list of free parameters in our chemical evolution model} \label{table:para_list}
\begin{tabular}{c|l} \hline \hline
$\rb$ & Break radius in the radial profile of gas infall \\
$\hrino$ & Scale length of infalling gas at the inside of $\rb$ \\
$\hrint$ & Scale length of infalling gas at the outside of $\rb$ \\
$\tauinz$ & Time scale of gas infall at $R$ = 0 kpc \\
$\tauine$ & Time scale of gas infall at $R$ = 8 kpc \\
$\alpha$ & Power-law index characterizing the radial \\ 
$ $ & dependence of the time scale of gas infall \\
$\Lambda_0$ & Mass loading factor at $R$ = 0 kpc \\
$\Lambda_8$ & Mass loading factor at $R$ = 8 kpc \\
$\beta$ & Power-law index characterizing the radial \\
$ $ & dependence of mass loading factor \\
$\arm$ & Radial gradient of diffusion length of disk stars \\
$ $ & by radial migration \\
$\brm$ & Diffusion length of disk stars by radial migration \\
$ $ & at $R$ = 0 kpc \\
$\gamma$ & Power-law index characterizing the dependence \\
$ $ & of radial diffusion length on stellar age \\
$\kre$ & Parameter controlling the radial profile of \\
$ $ & gas re-accretion \\
$\fia$ & Parameter controlling the rate of Type Ia supernova \\ \hline
\end{tabular}
\end{table}

Our chemical evolution model described above contains 14 free parameters, which are summarized in Table \ref{table:para_list}. In this study, we adopt the MCMC method (Metropolis et al. 1953; Hastings 1970) to obtain the best set of these free parameters, which reproduce the radial profile of star, gas, [O/H] and [Fe/H] of the ISM in the Galactic disk. In this model, following the results of several previous works, we adopt the Galactic stellar disk with the scale length of 2.3 kpc and the stellar density at $R$ = 8 kpc of 35 $\mr{M_\odot} \ \mr{pc}^{-2}$ (Flynn et al. 2006). For the total gas density profile, consisting of HI and H$_2$, in the Galactic disk, we adopt the work of Kubryk et al. (2015a) shown in their Figure A.2, and the radial profile of [O/H] and [Fe/H] in the disk is taken from the observation of Cepheids in Luck \& Lambert (2011). These observational constraints are necessary to be sure that our chemical evolution models are appropriate in comparison of the present Galactic stellar disk.


\begin{figure*}
\begin{center}
\includegraphics[width=16cm,height=12cm]{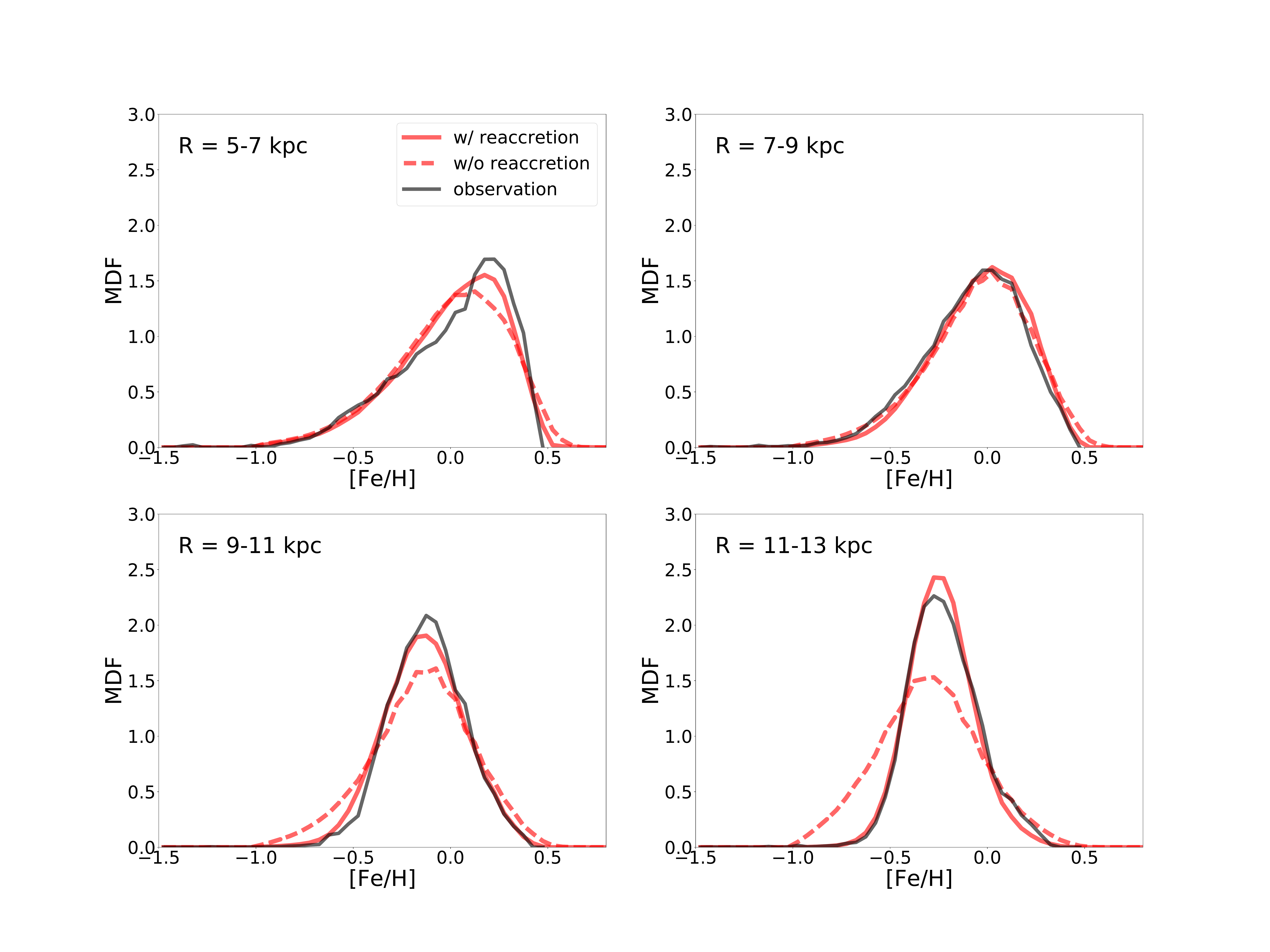}
\end{center}
\caption{The black lines represent the MDFs of the Galactic disk stars observed by the APOGEE survey (Hayden et al. 2015). The red solid and dashed lines are the MDFs obtained in our best fit model with and without the re-accretion of outflowing gas, respectively. The observed radial range of each MDF is denoted on the upper-left of each panel.}
\label{fig:mdf} 
\end{figure*}


\begin{figure*}
\begin{center}
\includegraphics[width=16cm,height=12cm]{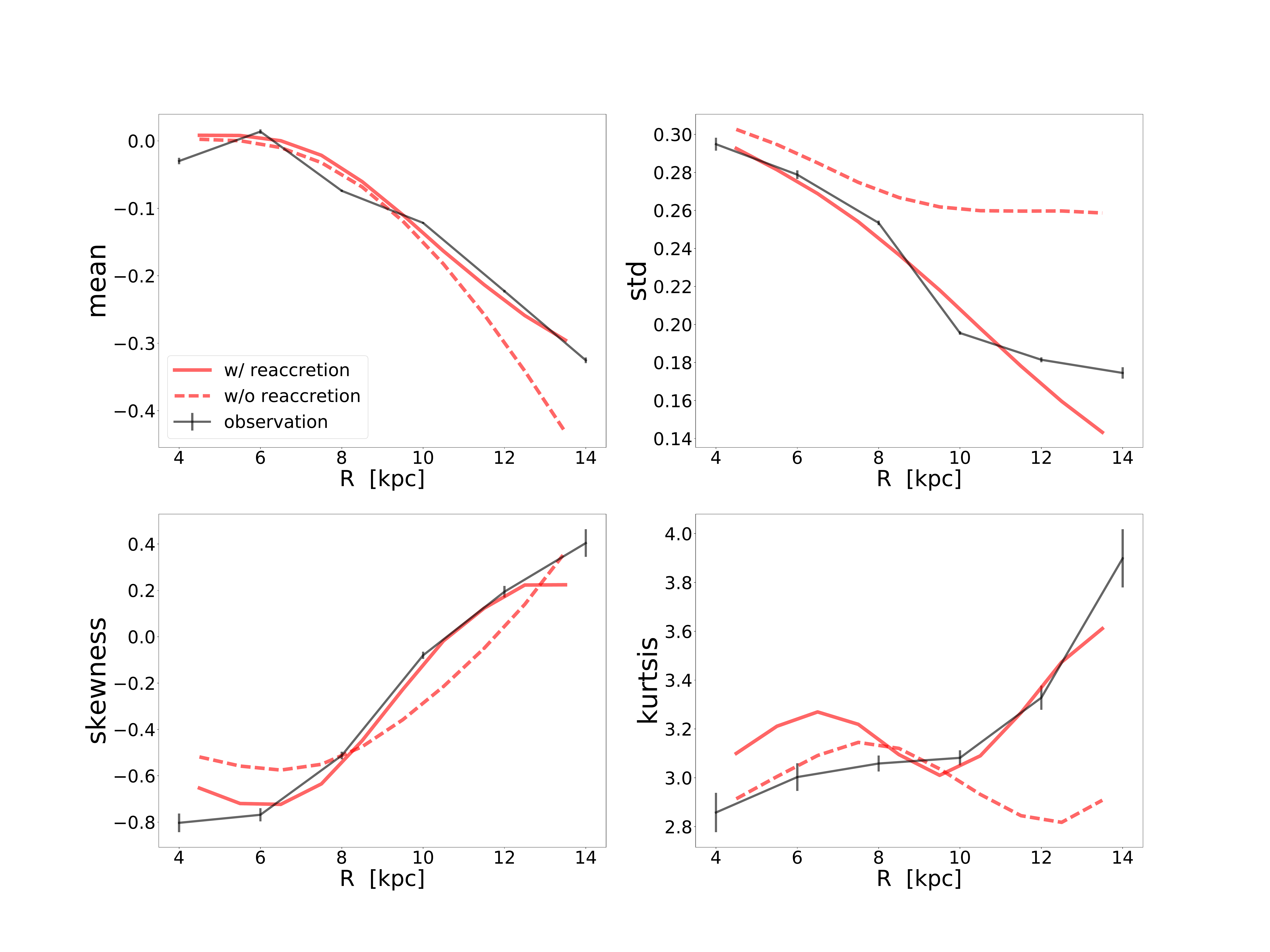}
\end{center}
\caption{The mean (upper left), standard deviation (upper right), skewness (lower left), and kurtosis (lower right panel) of the MDFs of disk stars as a function of $R$. The black lines represent the results for the observation in the APOGEE survey (Hayden et al. 2015), and their error bars correspond to the standard errors evaluated from the number of sample stars at the radius. The red solid and dashed lines are our best fit model with and without the re-accretion of outflowing gas, respectively.}
\label{fig:mom} 
\end{figure*}

Additionally, we also adopt the radial dependence of the MDFs of disk stars observed in APOGEE survey (Hayden et al. 2015) as the observational constraints to determine the model parameters in our MCMC procedure. The black lines of Figure \ref{fig:mdf} show the radial dependence of the MDF for the APOGEE red-giant star sample locating at $|Z|$ $<$ 2 kpc. We obtain these MDFs based on Figure 5 of Hayden et al. (2015), which shows the MDFs at different three ranges of $|Z|$, and their Table 1, which summarizes the number of sample stars at each range. It is clear that the properties of the MDFs, such as their peak and high and low-metallicity tales, significantly depend on the radii. Here, to quantify this radial change of the observed MDFs, we explore the mean, standard deviation, skewness, and kurtosis of the MDF at each radius as shown with the black lines of Figure \ref{fig:mom}. Each moment given in this figure is calculated only for $\feh > -1$ to exclude the contamination of metal-poor halo stars. The error bar of each plot represents the standard error evaluated from the number of sample stars at the radius. Figure \ref{fig:mom} shows that the mean and standard deviation decrease with increasing $R$, whereas the skewness increases from $\simeq -1$ at $R$ = 4 kpc to $\gtrsim$ 0 at $R$ = 12 kpc, which accompanies the growth of the high-metallicity tail in the MDF with increasing $R$. The kurtosis is mostly constant with $R$, but rapidly increases in the outer disk region, $R > 12$ kpc. In our MCMC procedure, we use these radial profile of the moments of the MDFs as the observational constraints. The shape of MDFs greatly reflects the star formation and chemical evolution histories, and thereby tells us useful information about the processes of gas infall, outflow, re-accretion, and radial migration of disk stars, as discussed in Section 4.

We use these constraints to obtain the posterior probability distributions and the best values of the 14 free parameters in the MCMC procedure. In each MCMC step, for a parameter set, we evaluate the value of the likelihood by investigating the difference of these constraints between our model prediction and the observation. For the calculation of the likelihood, we use the values of stellar mass, gas mass, [O/H] and [Fe/H] of the ISM at 5 points in the radial range of $R$ = 2-14 with the radial interval of 3 kpc, and those of the four moments of the MDF at 5 points in the radial range of $R$ = 5-13 kpc, where the MDFs are well observed by APOGEE, with the radial interval of 2 kpc. Thus we attempt to obtain the evolution history of the MW, which reproduces the observed MDF for the Galactic disk stars.

We mention here that red giant stars are a biased sample against older populations, and therefore the MDF in Figure \ref{fig:mdf} does not necessarily trace the true MDF of the underlying disk stars. To quantify this bias, Hayden et al. (2015) investigated the effect of sampling bias on the observed MDF based on the mock sampling of the APOGEE survey, and found that the systematic difference between the observed and the true MDFs is sufficiently small at least at $R >$  3 kpc. Thus, our fitting to the observed MDF in the radial range of $R$ = 5-13 kpc is justified to study the formation history of the Galactic stellar disk. 

Additionally, it may be worth noting that other observational properties of the Galactic disk, which we do not consider as constraints in our MCMC procedure, such as $\afe$-$\feh$ or age-$\feh$ relations, are also important clues to disentangle the formation history of the MW. In this work, we focus on the MDFs of disk stars, because they are very convenient to quantitatively compare model results with observational data via the values of their moments. In Section 5, we assess the significance of our model in terms of some other observational properties of the Galactic disk, and find our model can interpret these properties generally, but not completely. Therefore, constructing more sophisticated modeling, which incorporates other important observables as additional constraints, will be important as future work.

\begin{threeparttable}
\begin{center}
\caption{The results of the estimation of free parameters \label{table:result}}
\begin{tabular}{c|rrrr} \hline \hline
 & best\tnote{a} & 16 \%\tnote{b} & 50 \%\tnote{c} & 84 \%\tnote{d} \\ \hline
$\rb$ [kpc] & 5.66 & 5.13 & 5.59 & 6.18 \\
$\hrino$ [kpc] & 0.74 & 0.64 & 0.72 & 0.81 \\
$\hrint$ [kpc] & 3.75 & 3.44 & 4.17 & 5.38 \\
$\tauinz$ [Gyr] & 0.69 & 0.33 & 0.72 & 1.13 \\
$\tauine$ [Gyr] & 6.77 & 5.79 & 7.49 & 9.76 \\
$\alpha$ & 1.22 & 1.05 & 1.32 & 1.71 \\
$\Lambda_0$ & 0.26 & 0.06 & 0.12 & 0.26 \\
$\Lambda_8$ & 0.05 & 0.05 & 0.08 & 0.13 \\
$\beta$ & 0.16 & 0.14 & 0.5 & 1.86 \\
$\arm$ & 0.13 & 0.06 & 0.11 & 0.14 \\
$\brm$ [kpc] & 1.67 & 1.66 & 1.92 & 2.2 \\
$\gamma$ & 0.24 & 0.18 & 0.22 & 0.27 \\
$\kre$ & 0.25 & 0.17 & 0.24 & 0.31 \\
$\fia$ & 4.97 & 4.41 & 4.82 & 5.23 \\ \hline
\end{tabular}
\item[a] The best value obtained from the MCMC estimation.
\item[b] The 16 percentile value obtained from the MCMC estimation.
\item[c] The 50 percentile value obtained from the MCMC estimation.
\item[d] The 84 percentile value obtained from the MCMC estimation.
\end{center}
\end{threeparttable}

\section{FITTING RESULTS} \label{sec:result}

\begin{figure*}
\begin{center}
\includegraphics[width=18cm,height=18cm]{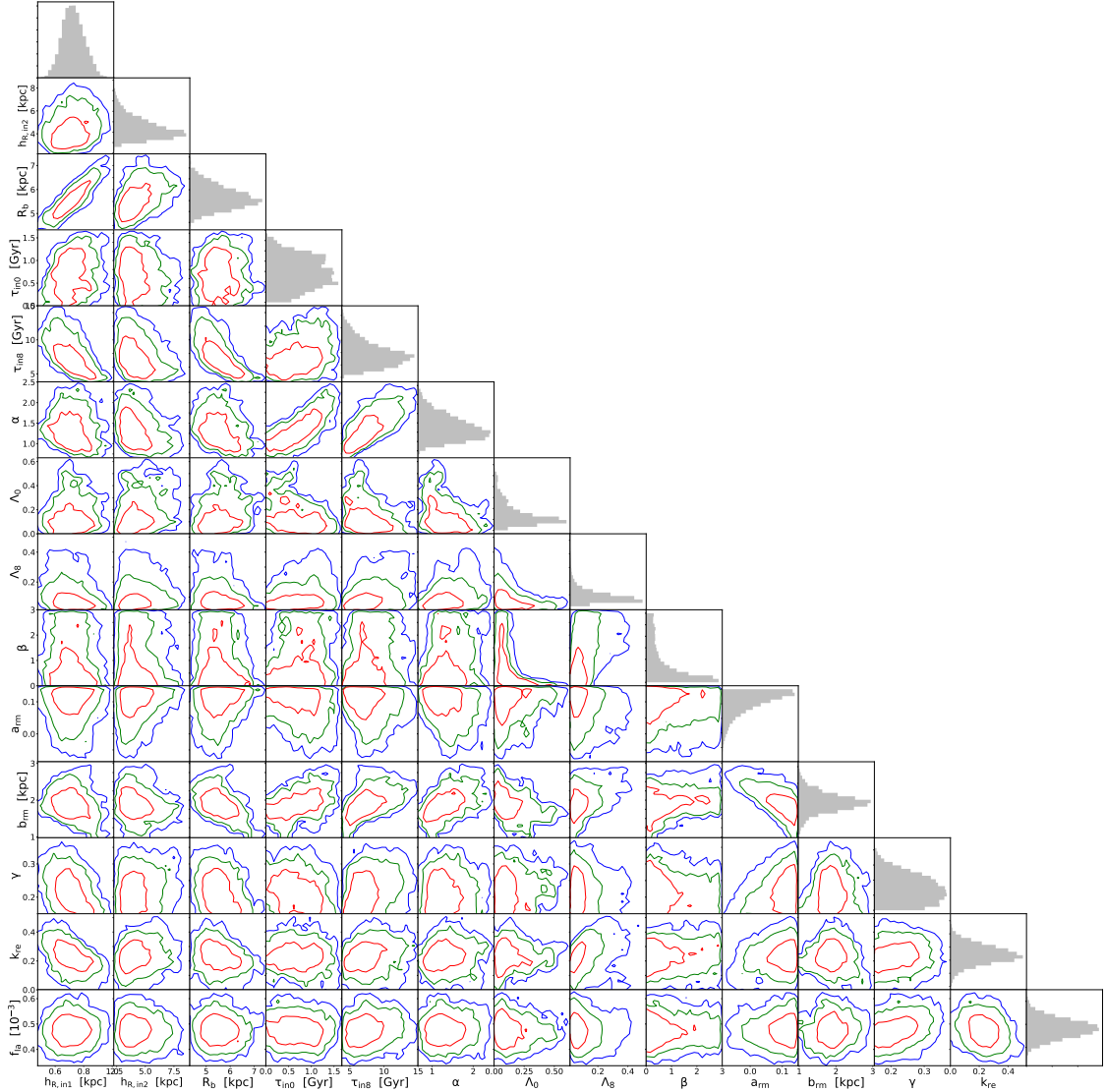}
\end{center}
\caption{The diagonal and off-diagonal panels show the 1D and 2D posterior probability distributions of the 14 parameters, respectively. The red, green and blue contours in each off-diagonal panel represent the 1, 2 and 3 $\sigma$ regions of the posterior probability distributions, respectively. }
\label{fig:mc} 
\end{figure*}

In our MCMC procedure, we carry out 5 MCMC chains starting from different parameter sets. Each chain consists of 200,000 iterations, and by compiling the later 100,000 iterations in all chains, we get the posterior probability distribution for each parameter. Figure \ref{fig:mc} shows the posterior probability distributions of the 14 parameters. We find from this figure that the MCMC chains for all parameters converge successfully. The results of this estimation of the 14 parameters are summarized in Table \ref{table:result}. 


\begin{figure*}
\begin{center}
\includegraphics[width=15cm,height=12cm]{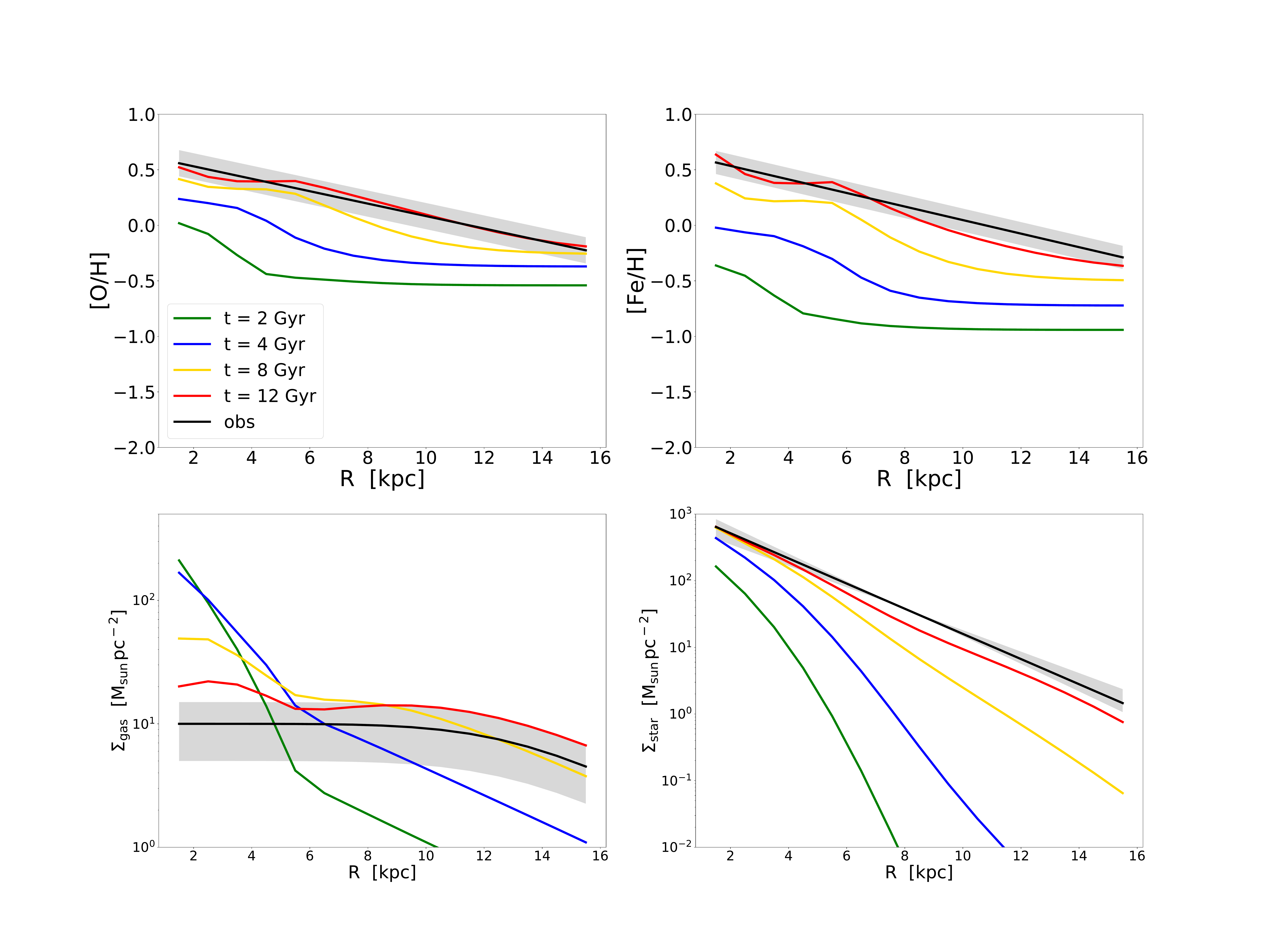}
\end{center}
\caption{The radial profiles of surface densities for [O/H], [Fe/H], gas, and stars obtained in the best fit model. The green, blue, yellow, and red lines show the results at $t$ = 2, 4, 8, and 12 Gyr, respectively. The black lines are the observed radial profiles, where shaded regions correspond to observational errors.}
\label{fig:rp} 
\end{figure*}

We here present the result of the model calculation based on the set of best-fit parameters. Figure \ref{fig:rp} shows the time evolution of the radial profiles of gas, star, [O/H] and [Fe/H] of the ISM. The green, blue, yellow and red lines in each panel represent the results at $t$ =  2, 4, 8, and 12 Gyr, respectively. These results imply that the structural and chemical evolutions have proceeded faster in the inner disk regions, corresponding to the inside-out formation scenario. Additionally, in each panel of Figure \ref{fig:rp} we show the observed profiles and their observational errors with the black lines and shaded regions, respectively. From comparison between the red and black lines with shaded regions in these figures, we find that our model can reproduce reasonably well the present radial profiles of chemical abundances of ISM in the Galactic disk. The gas and stellar profiles calculated here can be approximately fitted to the observed ones, but slightly over- and under-estimated, respectively. These slight differences between our fitting and the observation may suggest the necessity of some mass loss via gas outflow, which can decrease and increase gas and stellar mass densities at the fixed chemical abundances, respectively, although in this work we assume the recycling of all outflowing gas in order to reduce the number of free parameters.


The MDFs at the four radial ranges and the radial profiles of their mean, standard deviation, skewness, and kurtosis, reproduced by the best-fit model are shown with the red solid lines of Figure \ref{fig:mdf} and \ref{fig:mom}, respectively. We find from these two figures that our best fit model can generally reproduce the observed MDFs of the Galactic disk stars. Although the fitting to the value of kurtosis around $R$ = 6 kpc is not so well, the MDF as a whole at this radius seems to be reproduced reasonably well. Therefore, we consider that our fitting to the moments of the MDFs is carried out successfully.

Thus our chemical evolution models are appropriate in comparison of the Galactic stellar disk. In the next section, we show the detailed formation history of the MW predicted by the best fit model and present the implications for the gas infall, gas re-accretion and radial migration processes in the evolution of the MW stellar disk.

\section{FORMATION HISTORY OF THE MILKY WAY REPRODUCED BY MODEL CALCULATION} \label{sec:discussion}

\subsection{Star Formation History}

\begin{figure}
\begin{center}
\includegraphics[width=9cm,height=10cm]{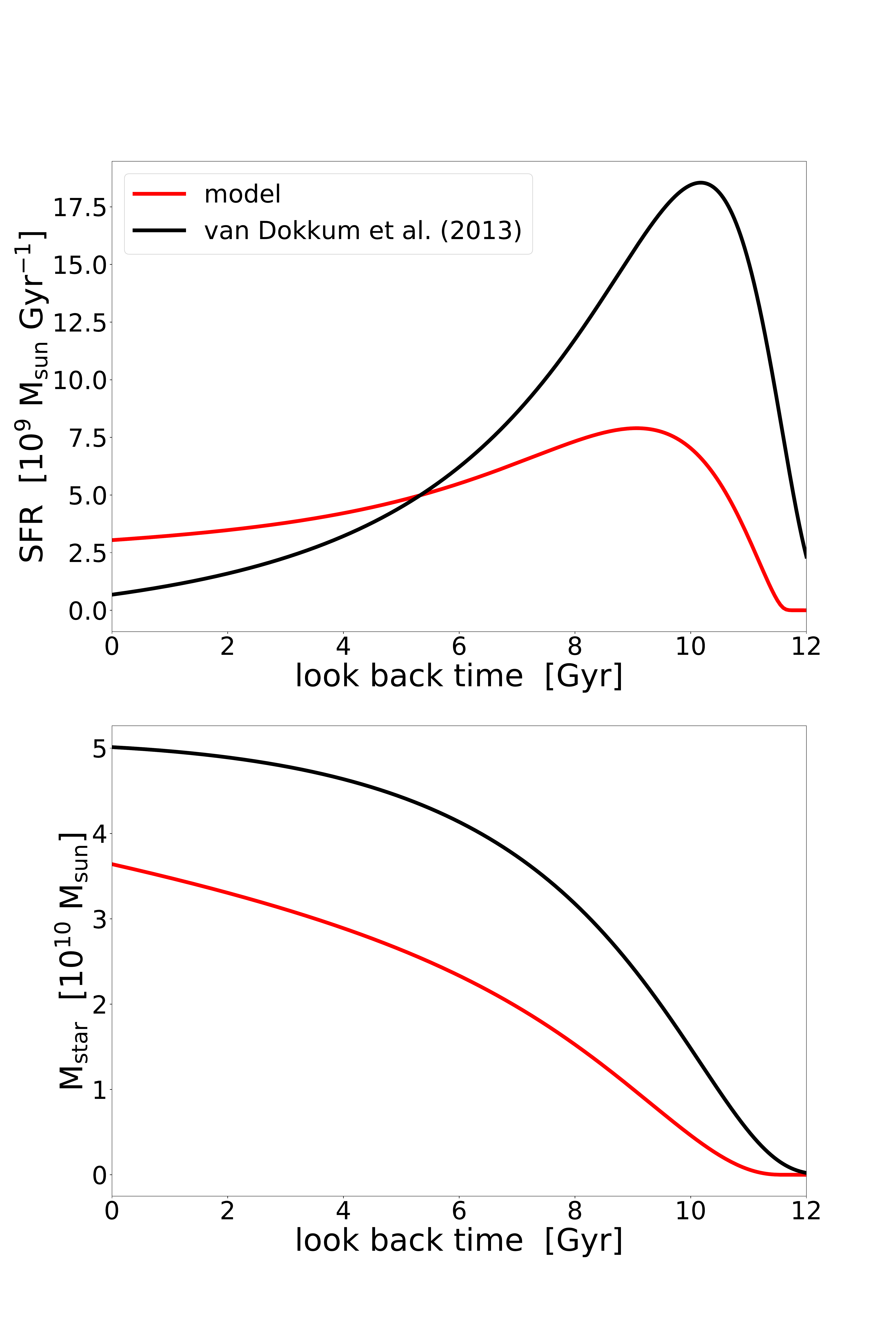}
\end{center}
\caption{The time evolution of the total star formation rate and the stellar mass in the galaxy are shown in the upper and lower panels, respectively. The red and black lines represent the results of our model calculation and the observation for the MW like galaxies by van Dokkum et al. (2013), respectively.}
\label{fig:sfr} 
\end{figure}

The red lines in upper and lower panels in Figure \ref{fig:sfr} represent the model results of the time evolution of total star formation rate and stellar mass, respectively. For comparison, the star formation history of the MW-like galaxies provided by van Dokkum et al. (2013) based on the abundance matching method are also shown with the black lines in these figures. While their present stellar mass is higher than that of our model by $\sim 10^{10} \ \msun$, this difference may be understood because the stellar mass from a bulge component is not included in our model. According to our model calculation, the star formation rate became highest at the look back time around 9 Gyr and more than half of the present stellar mass has already formed by the look back time of 6 Gyr, while the star formation activity at the early disk formation phase is much lower than that for van Dokkum et al. (2013). The star formation history in our model is very similar to that for the disk region provided by the model of Kubryk et al. (2015a).


\begin{figure}
\begin{center}
\includegraphics[width=9cm,height=6cm]{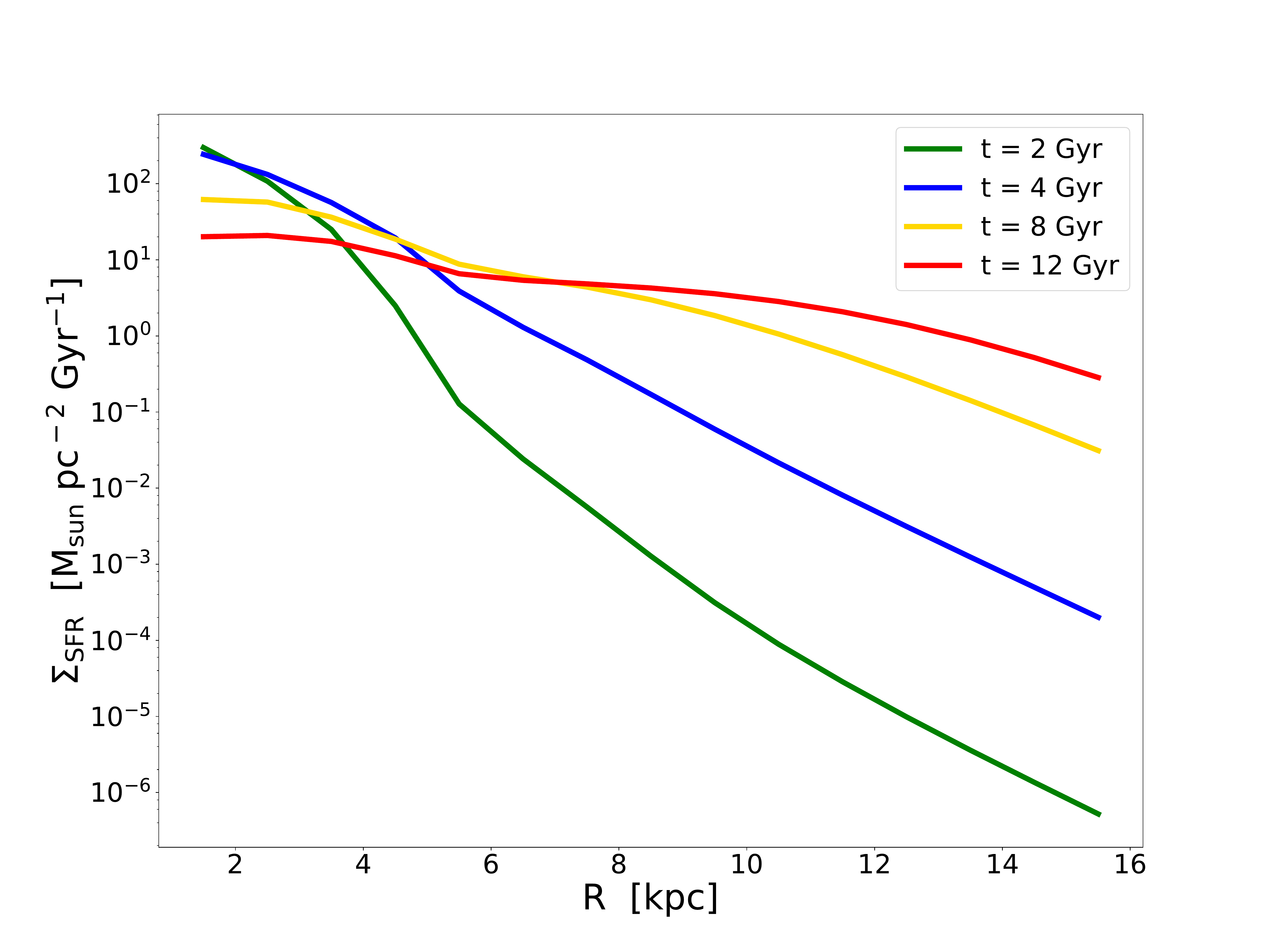}
\end{center}
\caption{The radial profiles of $\sgmsf$ as a function of $t$ obtained in the best fit model. The green, blue, yellow, and red lines show the results at $t$ = 2, 4, 8, and 12 Gyr, respectively.}
\label{fig:sgmsf} 
\end{figure}

The time evolution of the radial profile of $\sgmsf$ is depicted in Figure \ref{fig:sgmsf}. This figure shows that in the inner disk regions the star formation activity is significantly high at the early disk formation phase, and declines with increasing time. In contrast, the star formation rate in the outer disk regions is lower than that in the inner ones at all epochs, but gradually increases with increasing time. This star formation history corresponds to the inside-out disk evolution as shown in Figure \ref{fig:rp}. In the following subsections, we investigate the detailed properties of each process, which can significantly affect the star formation history.

\subsection{Gas Infall History}

\begin{figure}
\begin{center}
\includegraphics[width=9cm,height=6cm]{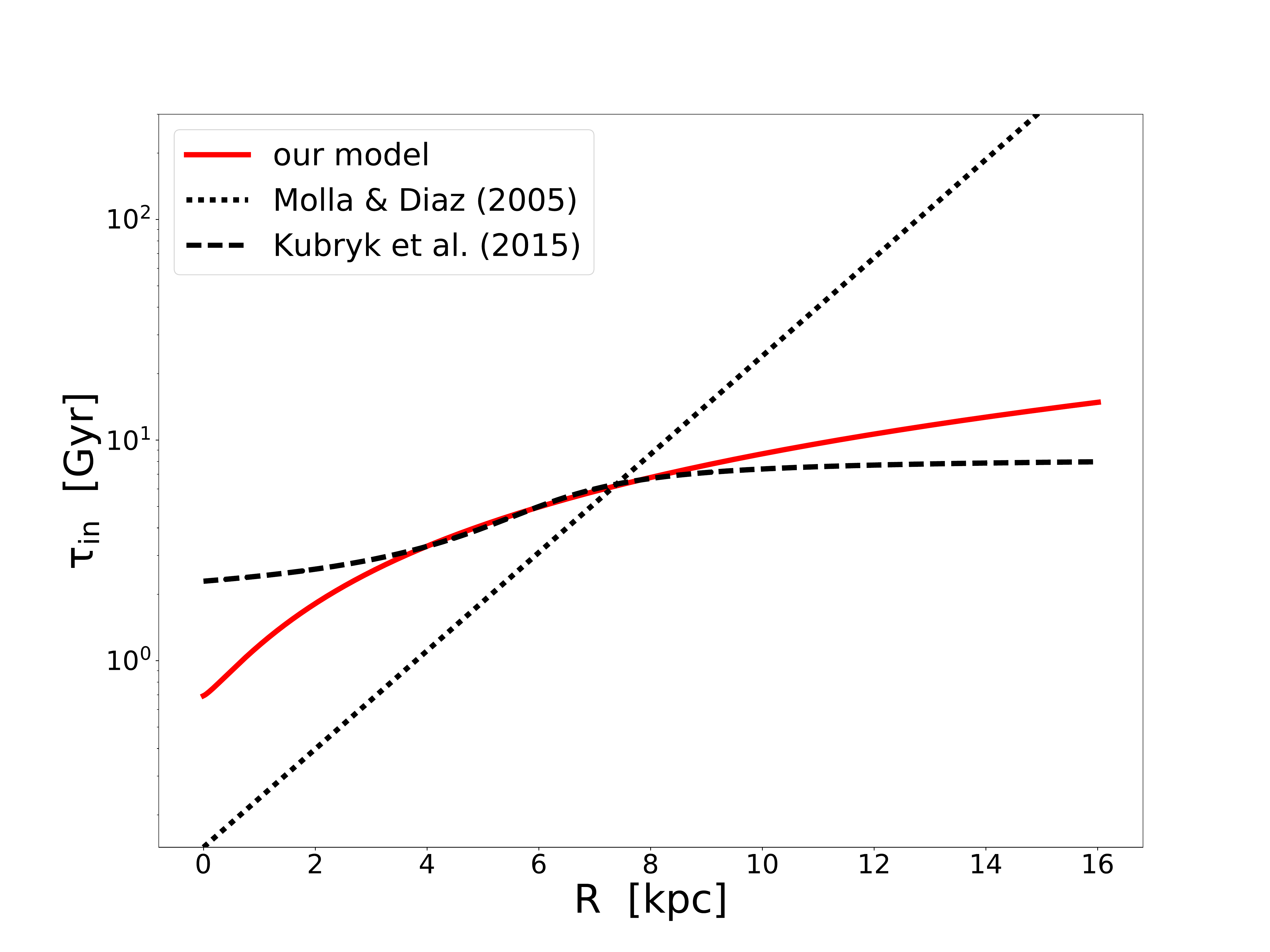}
\end{center}
\caption{The red line shows the radial profile of $\tauin$ in our best fit model. For comparison, the radial profiles of $\tauin$, adopted in the model of Molla \& Diaz (2005) and Kubryk et al. (2015) are described with the black dotted and dashed lines, respectively.}
\label{fig:tauin} 
\end{figure}

The red line in Figure \ref{fig:tauin} presents the radial profiles of time scale of gas infall based on the best fit parameters. For comparison, we also plot the time scale of gas infall adopted in previous models of Moll{\'a} \& D{\'{\i}}az (2005) and Kubryk et al. (2015a) with black dotted and dashed lines, respectively. Figure \ref{fig:tauin} shows that our time scales of gas infall are around 1 Gyr and the Hubble time in the central and outer disk regions, respectively, thereby suggesting that our model result is an intermediate case between Moll{\'a} \& D{\'{\i}}az (2005) and Kubryk et al. (2015a).  This shorter $\tauin$ in the inner disk region is simply understood with a shorter cooling and collapsing time in an inner gas halo, in which the gas density is higher.  In our MCMC fitting, such an inside-out gas infall history is naturally preferred because the gas to stellar mass ratio is larger in the outer disk regions in the MW, implying that the disk evolution has proceeded faster in the inner regions. 


\begin{figure}
\begin{center}
\includegraphics[width=9cm,height=6cm]{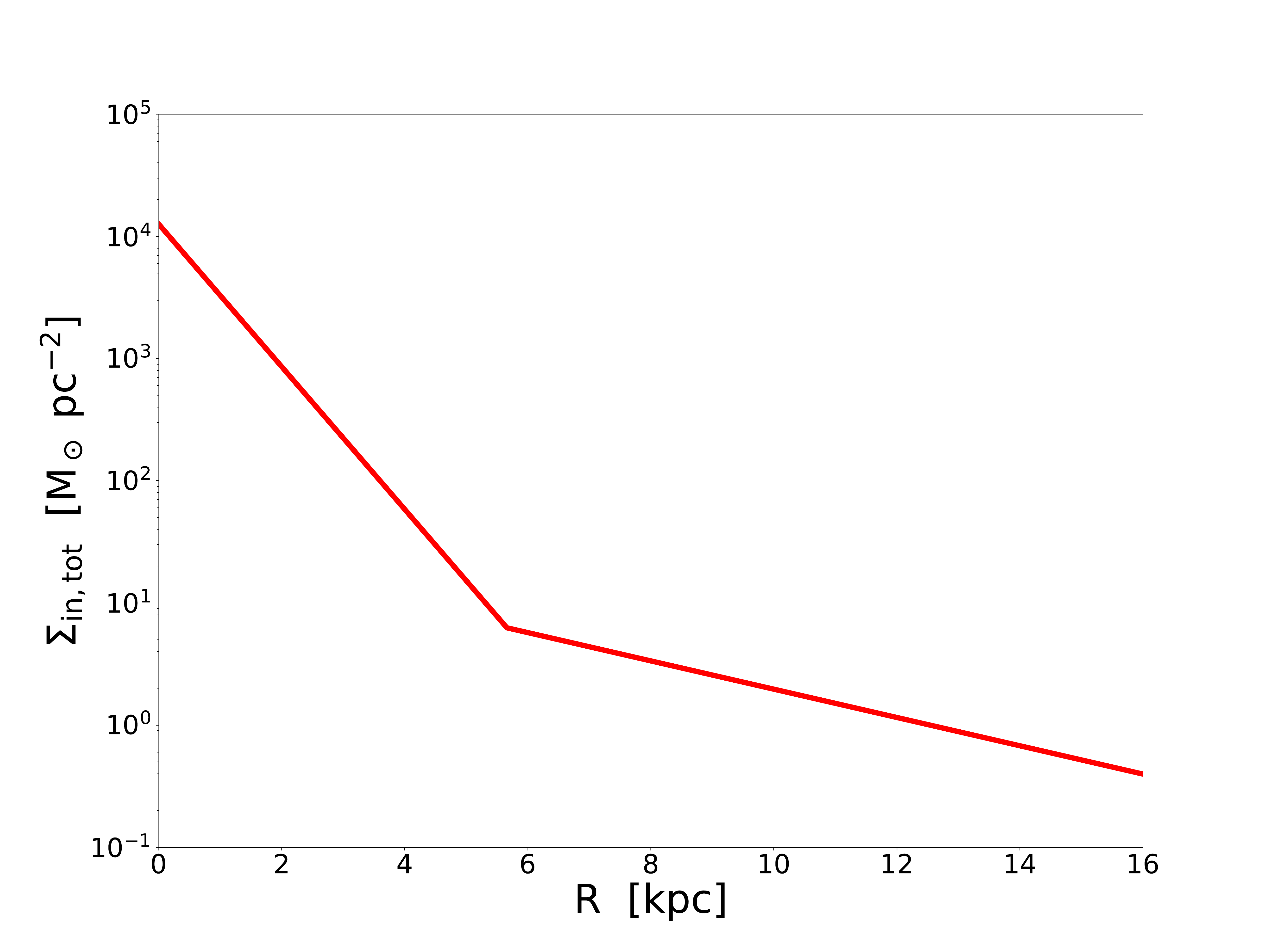}
\end{center}
\caption{The radial profile for the surface density of the total mass of gas inflow in the best fit model.}
\label{fig:total_sgmin} 
\end{figure}

We show the surface density of the total mass of gas infall as a function of $R$ in Figure \ref{fig:total_sgmin}. We find that the radial profile of gas infall is clearly up-bending. An up-bending profile, which is more centrally concentrated than the present stellar density profile, is favored in our MCMC procedure, because an original stellar density profile must be radially extended via stellar radial migration process, as will be noted in Section 4.4 in detail. 


Here, to understand the origin of such an up-bending profile, we refer a theoretical prediction by van den Bosch et al. (2001). The prediction is based on an assumption that gas in a halo follows the same specific angular momentum distribution as the dark matter, as produced in N-body simulations of structure formation, and all the gas cool and form the disk by conserving the specific angular momentum distribution completely. The radial profile of gas infall from this simple prediction is a double power law rather than a single exponential, and it is generally similar to our up-bending profile. This fact suggests that such an up-bending profile of gas infall results from the original distribution of angular momentum of gas accreting onto the galactic disk plane. However, we note here that the above assumption about the angular momentum conservation of gas is somewhat simplistic because gas can cool and form fine structures through energy dissipation process, and consequently the gravitational torque to gas component is much different from that to the dark mattar component (Danovich et al. 2015). Therefore the origin of the up-bending profile of gas infall should be investigated more strictly based on hydrodynamical simulations.

\begin{figure}
\begin{center}
\includegraphics[width=9cm,height=6cm]{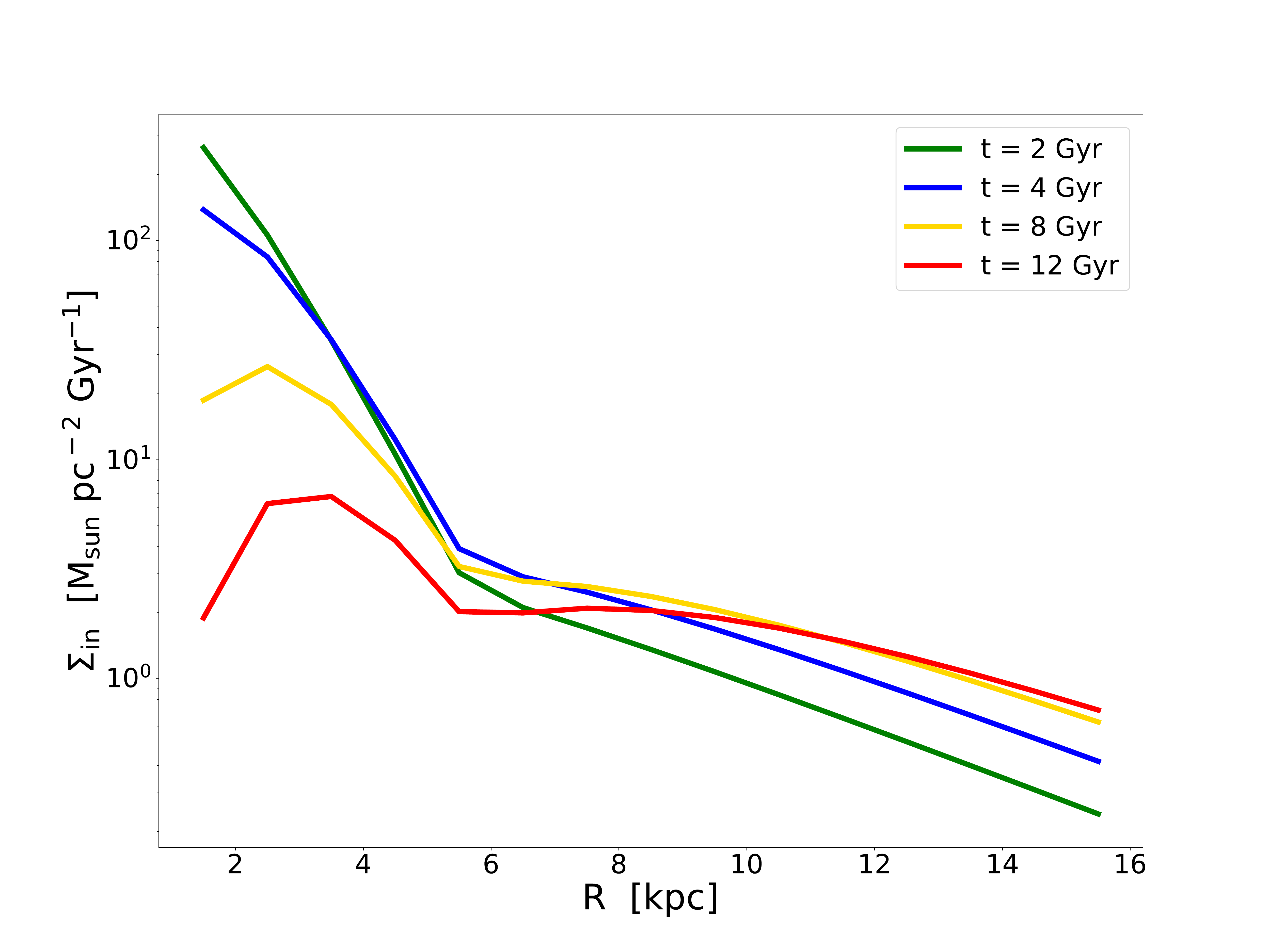}
\end{center}
\caption{The radial profile of $\sgmin$ as a function of $t$ in the best fit model. The green, blue, yellow, and red lines show the results at $t$ = 2, 4, 8, and 12 Gyr, respectively.}
\label{fig:sgmin} 
\end{figure}

The resulting gas infall history is shown as a function of $R$ and $t$ in Figure \ref{fig:sgmin}. As expected from the radial profile of $\tauin$ shown in Figure \ref{fig:tauin}, at the early disk formation phase a large amount of gas rapidly accretes onto the inner regions of $R \simeq$ 6 kpc, whereas in the outer regions gas stationary accretes over the whole disk formation phase. This infall history is a key ingredient leading to the inside-out star formation history described in the previous section.


\subsection{Importance of Re-accretion of Outflowing Gas }

\begin{figure}[h]
\begin{center}
\includegraphics[width=9cm,height=6cm]{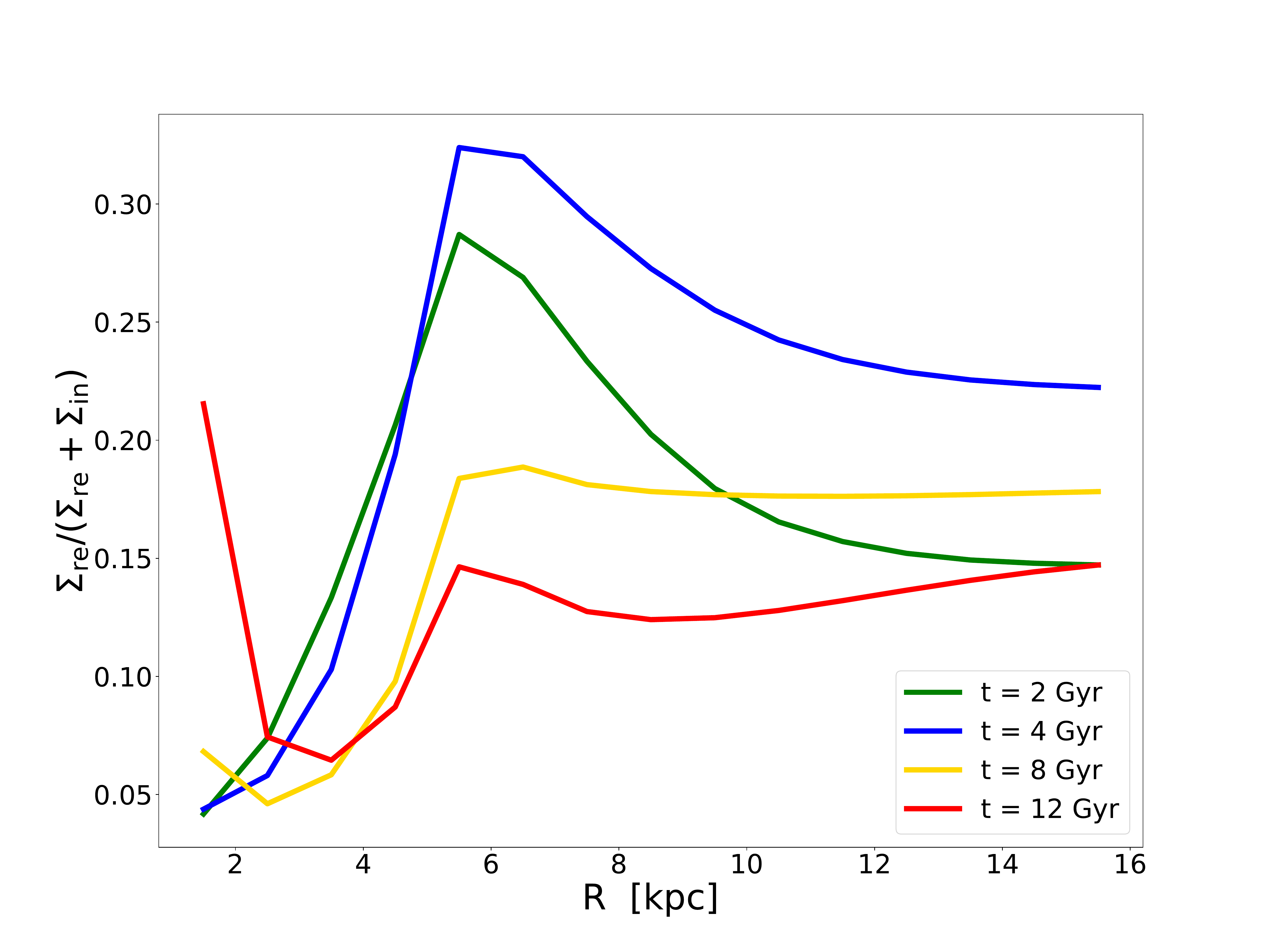}
\end{center}
\caption{The radial profile of $\sgmre/(\sgmre + \sgmin)$ as a function of $t$ in the best fit model. The green, blue, yellow, and red lines show the results at $t$ = 2, 4, 8, and 12 Gyr, respectively.}
\label{fig:reacc} 
\end{figure}

Our MCMC procedure gives the best fit solution of $\kre$ = \krebest, suggesting that gas ejected from the Galactic disk can fall back preferentially onto the outer disk regions, which is in good agreement with the prediction from the previous numerical simulation (Bekki et al. 2009). This re-accretion of once ejected gas is important as a gas supply mechanism into the Galactic disk. Figure \ref{fig:reacc} represents the ratio of $\sgmre$ to $(\sgmre + \sgmin)$, i.e., total mass of gas supply, as a function of $t$ and $R$. This shows that for the entire disk, the gas re-accretion can be a non-negligible contributor of gas supply at $t >$ 2 Gyr, i.e., after the early phase that the star formation activity is still confined at the inner disk region; at $t \sim$ 4 Gyr, $\sgmre$ can hold $ \gtrsim$ 20 \% of total gas supply. 

\begin{figure}[h]
\begin{center}
\includegraphics[width=9cm,height=6cm]{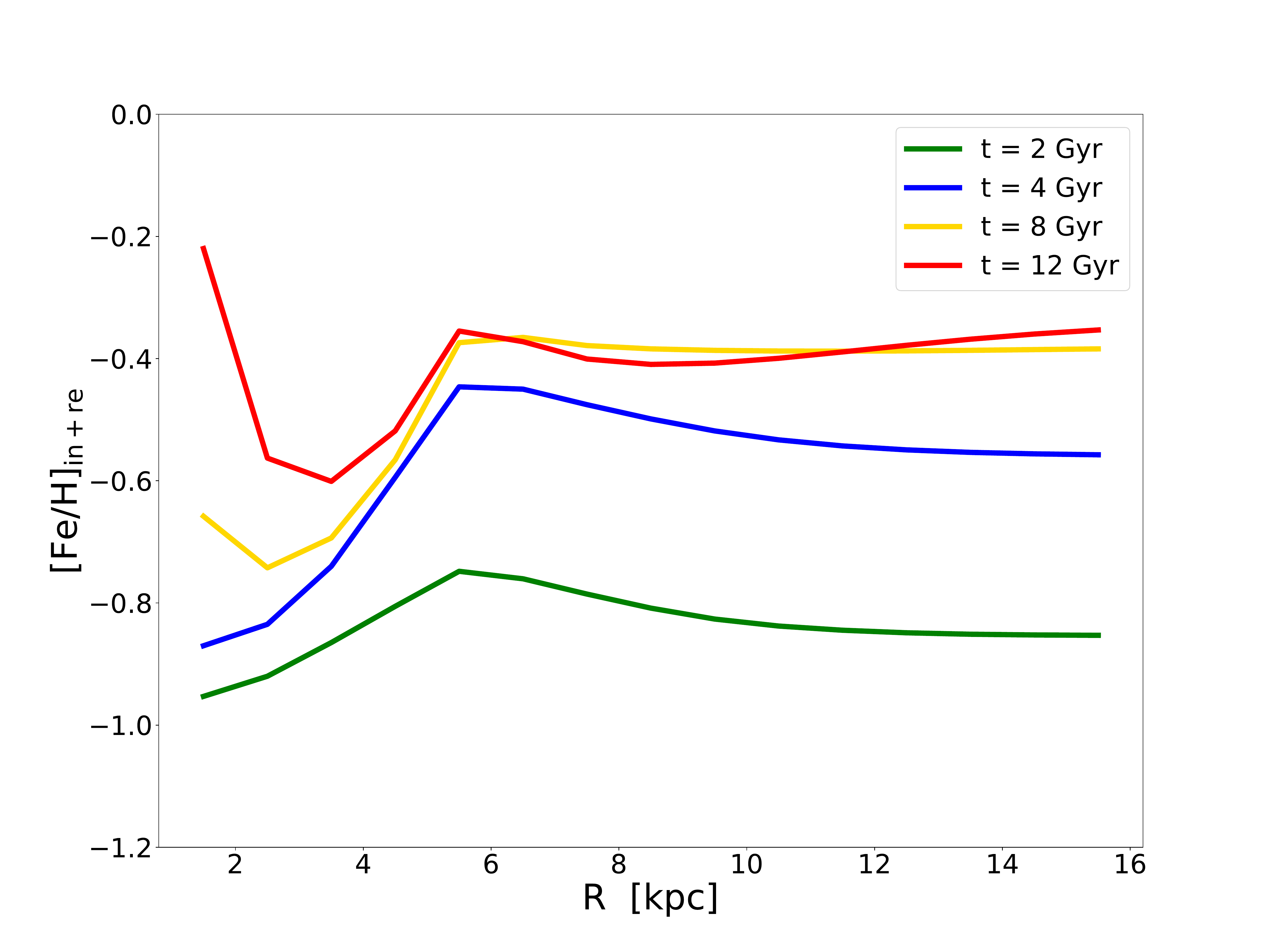}
\end{center}
\caption{The radial profile of $\mr{[Fe/H]_\mr{in+re}}$ as a function of $t$ in the best fit model. The green, blue, yellow, and red lines show the results at $t$ = 2, 4, 8, and 12 Gyr, respectively.}
\label{fig:reaccfe} 
\end{figure}

This re-accretion process of metal-enriched gas is found to be equally important for the chemical evolution of the Galactic disk. In Figure \ref{fig:reaccfe}, we show the metallicity of the mixed gas with the metal-poor infalling gas and metal-enriched re-accreting gas, $\mr{[Fe/H]_\mr{in+re}}$, as a function of $t$ and $R$. It is evident that $\mr{[Fe/H]_\mr{in+re}}$ tightly relates with $\sgmre/(\sgmre + \sgmin)$ shown in Figure \ref{fig:reacc}. In particular, in the outer disk regions of $R > 6$ kpc, $\mr{[Fe/H]_\mr{in+re}}$ has already increased to $\gtrsim -0.6$ for the first 4 Gyrs, because re-accretion of metal-enriched outflowing gas can contribute to gas supply onto the Galactic disk as well as metal-poor gas infall. Thus, our model calculation predicts that the outer regions in the Galactic disk have been strongly chemically polluted by the re-accreting metal-enriched gas, originally ejected from the inner disk regions. Such a chemical pollution by re-accreting gas can be seen in the galaxy formation simulation of Gibson et al. (2013). Also, the recent observations of nearby galaxies having extended gas disks suggest the necessity of such transportations of metal-enriched gas from inner to outer disk regions, probably via re-accretion of outflowing gas, to explain the metallicities of the ISM at the outskirts of the disk galaxies (Werk et al. 2010, 2011; Bresolin et al. 2012).

To understand the importance of such a re-accretion process in the formation history of the MW, we here present an additional MCMC experiment for the case without any re-accretion of outflowing gas. The red dashed lines in Figure \ref{fig:mdf} and \ref{fig:mom} are the same as the red solid ones in these figures, but for the best fit result without the effect of re-accretion of outflowing gas.  It is evident that for the case without gas re-accretion, significant metal-poor tales in the MDFs are formed especially in the outer disk regions, and consequently the standard deviation of the MDFs becomes much larger than the observed one. This result implies that the re-accretion of metal-enriched outflowing gas is a key ingredient to reproduce the observed MDFs of disk stars in the outer parts of the Galactic disk. 

The reason why gas re-accretion produces the narrower MDFs at the outer disk regions can be simply understood as follows. If gas re-accretion as shown in Figure \ref{fig:reacc} and \ref{fig:reaccfe} occurs, then in the outer disk regions having longer gas infall time scales than in the inner regions as shown in Figure  \ref{fig:tauin}, stars are formed from metal-enriched gas with $\feh \gtrsim -0.6$, supplied from re-accretion. As a result, the MDFs in the outer disk regions do not extend to $\feh < -0.7$. On the other hand, for the case without gas re-accretion, in the outer disk region stars are born from metal-poor gas and the observed small standard deviations of the MDFs are never reproduced. Thus, we conclude, based on the re-accretion of outflowing gas and the inside-out star formation history obtained from our model calculation, that the inner disk regions of the MW formed very rapidly from metal-poor gas, whereas the outer regions have been gradually constructed from metal-rich gas supplied from re-accretion of once ejected gas via galactic wind. We suggest that this formation history of the Galactic stellar disk is imprinted in the radial dependence of the low metallicity side of the MDFs of the disk stars.

We mention here that radial migration of disk stars is a secondary process for making the MDFs narrower in the outer disk regions, as evident from the fact that both of the two models depicted with the red solid and dashed lines in Figure \ref{fig:mdf} include the effect of radial migration. On the other hand, the formation of the high-metallicity tails of the MDFs are mainly responsible for stellar migration process as shown in the next subsection.

Additionally, we also note that in our chemical evolution model the effect of gas radial flow along a galactic disk, expected to result from gravitational interactions between gas and spiral/bar or giant molecular clouds (e.g., Yoshii \& Sommer-Larsen 1989), is not taken into account because of the large computational cost. However, such a radial flow of gas can transport metals from inner to outer disk regions, and thereby may provide a similar effect to the re-accretion of metal-enriched gas discussed above. We would like to consider this subject in our future work.


\subsection{Radial Migration History}

\begin{figure}[h]
\begin{center}
\includegraphics[width=9cm,height=6cm]{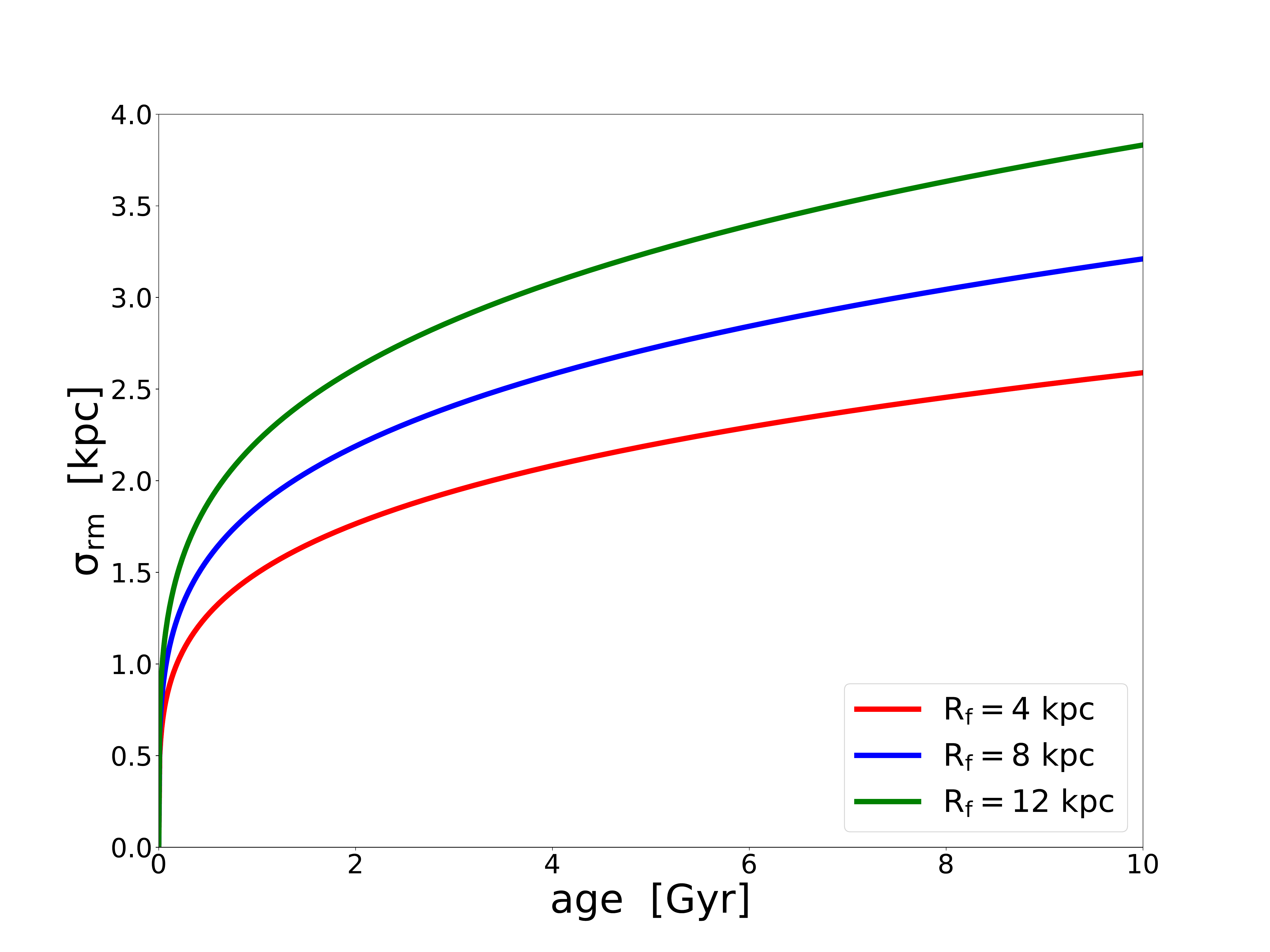}
\end{center}
\caption{The dependence of $\srm$ on stellar age in the best fit model. The red, blue, and green lines are $\srm$ for $\rf$ = 4, 8, and 12 kpc, respectively.}
\label{fig:srm} 
\end{figure}

Our MCMC procedure determines the best fit parameters for the radial migration history as ($\arm$, $\brm$, $\gamma$) = (\armbest, \brmbest \ kpc, \gnmbest). The evolution of $\srm$ based on the best fit parameters is demonstrated in Figure \ref{fig:srm}. Here, we show the cases for $\rf$ = 4, 8 and 12 kpc with red, blue, and green lines, respectively, and find that the diffusion length of radial migration is smaller for disk stars born in the inner disk regions. Our MCMC fitting prefers such a smaller diffusion length in the inner disk regions to reproduce the low-metallicity tail of the MDFs. As noted in the previous section, in our model with gas re-accretion, the formation of metal-poor stars with $\feh \sim -1$ are confined in the inner disk regions, $R <$ 4 kpc. Then, much larger diffusion lengths in the inner disk regions, which lead to too significant net outward transfer of disk stars, are disfavored, because they make a significant low-metallicity tail extending to $\feh \sim -1$ even in the outer disk regions. However, we note that the numerical simulation for a MW-like galaxy by Kubryk et al. (2013) provides a negative $\arm$, corresponding to a smaller $\srm$ in larger radii, because an inner disk is more gravitationally unstable and consequently has more bar/spiral structures and GMCs, triggering the radial migration of disk stars. Therefore, we re-calculated with some fixed negative values of $\arm$, and found that our conclusion in this paper is not modified, at least for $\arm > -0.15$, which is much steeper than $\arm = -0.067$ obtained in Kubryk et al. (2013).

Additionally, Figure \ref{fig:srm} shows that the growth rate of $\srm$ rapidly decreases with increasing age. This evolution of $\srm$ implies that although young disk stars can effectively move radially due to interactions with bar/spiral structures and GMCs, the efficiency of radial migration of disk stars quickly declines with increasing stellar age because such interactions simultaneously disturb the dynamics of disk stars and weaken the gravitational connections with bar/spirals and GMCs (e.g., H{\"a}nninen \& Flynn 2002; De Simone et al. 2004; Aumer et al. 2016).

\begin{figure}[h]
\begin{center}
\includegraphics[width=9cm,height=6cm]{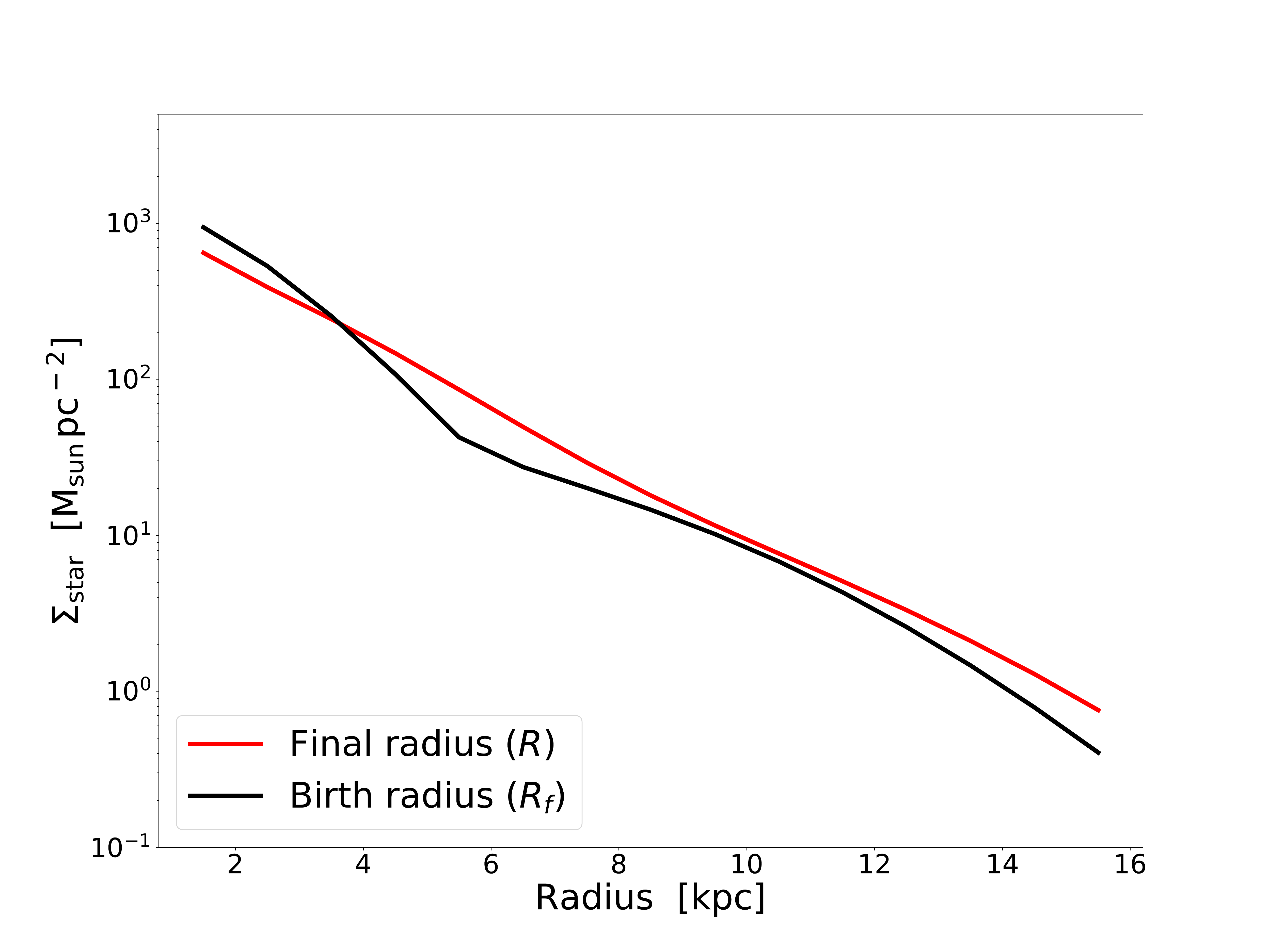}
\end{center}
\caption{The stellar density as a function of the birth and final radii, namely $\rf$ and $R$, respectively, in the best fit model.}
\label{fig:star_nrm} 
\end{figure}

We here investigate the influence of this radial migration history on the present stellar distribution in the disk. Figure \ref{fig:star_nrm} shows the stellar densities as a function of the birth and final radii, namely $\rf$ and $R$, with the red and black lines, respectively. The stellar density profile for $\rf$ is up-bending with the break around $\rf \sim$ 6 kpc as reflecting the radial profile of gas infall as shown in Figure \ref{fig:total_sgmin}, while that for $R$ is roughly described as an exponential profile with no clear break and is less centrally concentrated than for $\rf$. This difference of stellar density profiles for $\rf$ and $R$ suggests that radial migration process has occurred the net transport of disk stars from the inner high-density to the outer low-density disk regions.

\begin{figure}[h]
\begin{center}
\includegraphics[width=9cm,height=12cm]{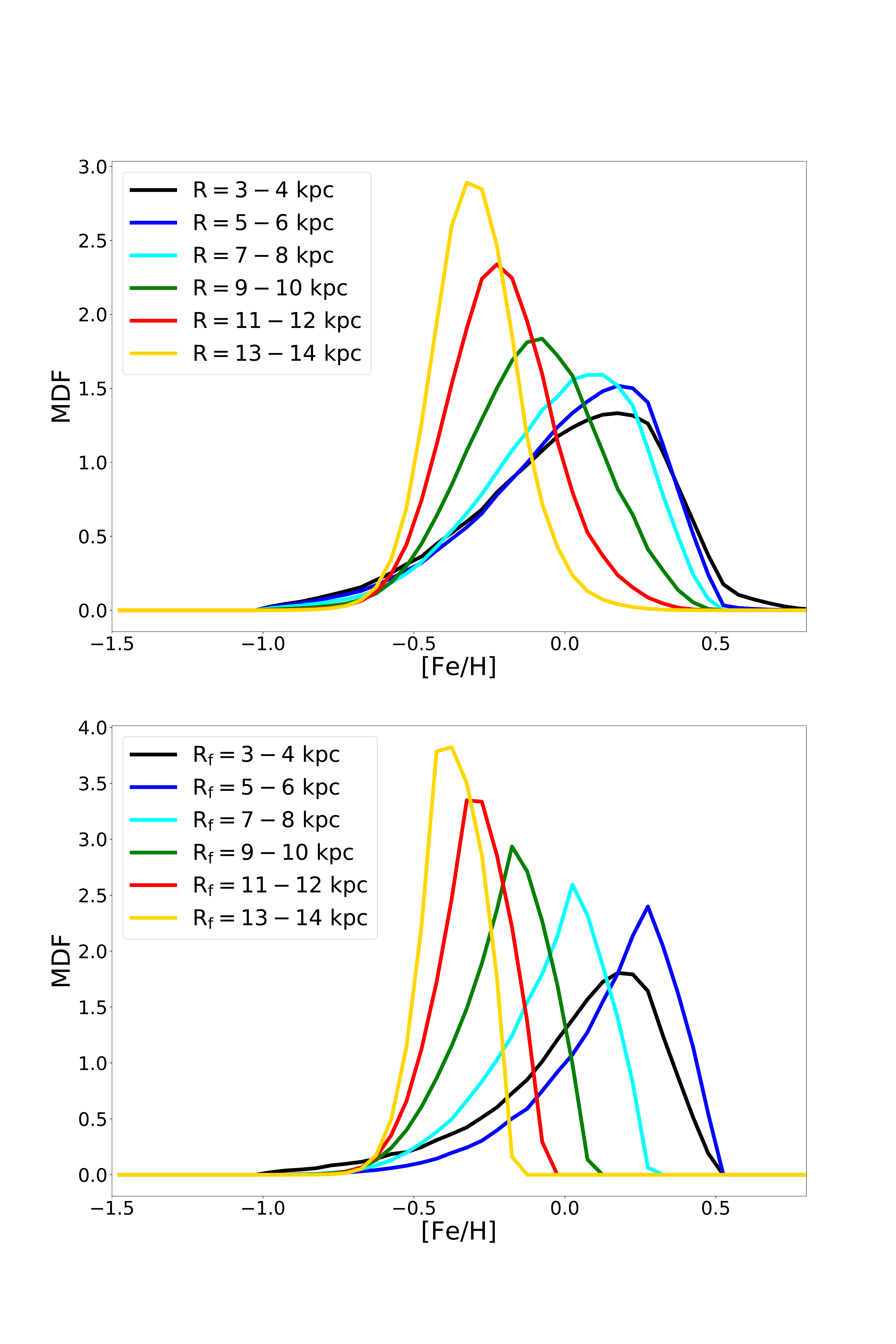}
\end{center}
\caption{The upper and lower panels show the MDF as a function of $R$ and $\rf$, respectively, in the best fit model.}
\label{fig:mdf_rf} 
\end{figure}

Moreover, to clarify the impact of radial migration process on the MDF of the disk stars, we show the MDF as a function of $R$ and $\rf$ in the top and bottom panels, respectively, in Figure \ref{fig:mdf_rf}. From these panels it is found that the MDFs observed in any $R$, especially in the outer disk regions, have more significant high-metallicity tails than the MDFs of disk stars with the same $\rf$. This implies that the observed high metallicity tails in MDFs in the outer disk regions are revealed due to the net radial transfer of metal-rich disk stars from the inner to outer disk regions. A similar suggestion was recently presented in Loebman et al. (2016) based on the numerical simulation for the formation of an isolated disk galaxy. Thus, the radial migration effects are necessary to interpret the observed radial change of the shape of MDFs.


\subsection{Short Summary} 

In this subsection, we briefly summarize the properties of the formation history of the MW disk obtained from our model calculation, and what observational information is important to constrain each process.

In our best fit model, gas infall occurs more rapidly onto the inner disk regions. Such an inside-out gas infall history is naturally favored to reproduce the larger gas to stellar mass ratio in the outer disk regions of the MW. The radial profile of gas infall is clearly up-bending as expected to reflect from the angular momentum distribution of infalling gas. The up-bending profile of gas infall is obtained to fit the observed stellar density profile by combining with the stellar radial migration effect. Thus, the gas-infall history is mainly constrained from the observed spatial distribution of gas and stars in the Galactic disk.

Gas outflow and subsequent gas re-accretion are important mechanisms, which can transport metals from the inner to outer disk regions, and therefore significantly affects the spatial distribution of heavy elements in the disk. Moreover, the combination of such a metal-enriched gas re-accretion and inside-out disk formation is an essential ingredient to reproduce the low-metallicity tails of the MDFs of disk stars. Namely, the inner disk regions of the MW formed rapidly from metal-poor gas, whereas the outer regions have been gradually constructed from metal-rich gas supplied from re-accretion, and consequently the MDFs become narrower in the outer disk regions as actually observed in the MW. In summary, the radial profiles of chemical abundance of the ISM and the standard deviation of the MDF are important observables to constraint the properties of gas re-accretion process.

Stellar radial migration leads to net transports of stars from the inner high-density to the outer low-density regions, and thereby can modify an originally up-bending stellar density profile to a less centrally concentrated one with no clear break. Additionally, the migration of metal-rich stars toward outer disk regions can form the significant high-metallicity tails of MDFs, as also suggested in Hayden et al. (2015) and Loebman et al. (2016). Therefore, the radial migration history is constrained mainly from the radial profile of the skewness of MDFs and also partly from the stellar density profile.

Thus, our new model calculation suggests that gas infall, re-accretion of outflowing gas, and stellar radial migration are essential processes in the formation of the MW stellar disk, and the property of each process can be generally constrained from individual observational clues in our MCMC procedure.

\section{OTHER OBSERVATIONAL PROPERTIES IN THE GALACTIC STELLAR DISK} \label{sec:other}

Here, to assess the significance of our model results, we discuss the other important observational properties in the Galactic stellar disk based on our best fit model.

\subsection{Stellar distribution on the $\afe$-$\feh$ plane}


\begin{figure*}
\begin{center}
\includegraphics[width=18cm,height=14cm]{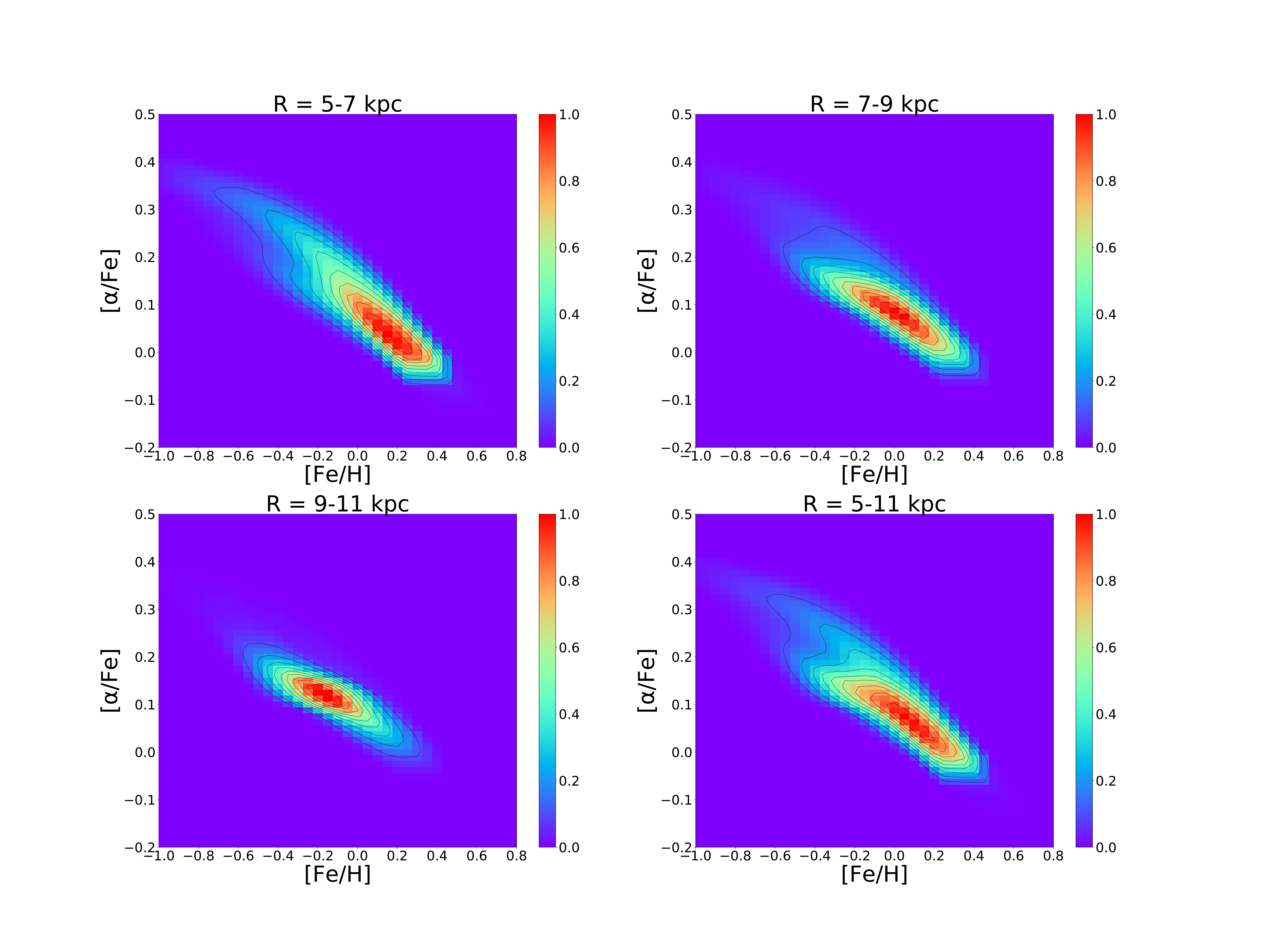}
\end{center}
\caption{The color maps represent the stellar distributions on the $\afe-\feh$ plane at $t$ = 12 Gyr, reproduced in the best fit model. The stellar distribution observed at the inner ($R$ = 5-7 kpc), solar neighborhood ($R$ = 7-9 kpc), the outer ($R$ = 9-11 kpc), and the summation of the three regions ($R$ = 5-11 kpc) are shown from the top-left to bottom-right panels, respectively.}
\label{fig:afe-feh_dist} 
\end{figure*}

Many of previous observations have reported that the distribution of the Galactic disk stars on the $\afe$-$\feh$ plane is bimodal: there are the high-$\afe$ and low-$\afe$ peaks in the stellar density at the $\afe$ ratio of  $\sim$ 0.2-0.3 and $\sim$ 0, respectively (e.g., Bensby et al. 2003; Lee et al. 2011; Adibekyan et al. 2012; Anders et al. 2014; Mikolaitis et al. 2014). Recently, by using the APOGEE red-giant sample, Hayden et al. (2015) investigated the dependence of such a bimodality on the distances from the Galactic center, $R$, and the disk plane, $|Z|$. They found that the high-$\afe$ peak is more significant than the low-$\afe$ one at smaller $R$ and larger $|Z|$, implying that the high-$\afe$ disk population is geometrically thicker and more concentrate than the low-$\afe$ one. Additionally, although the high-$\afe$ sequence does not strongly change along the disk plane, the metallicity of the low-$\afe$ peak clearly decreases with increasing $R$. These observed properties of the stellar distribution on the $\afe$-$\feh$ plane are closely associated with the formation history of the Galactic stellar disk. 

However, the origin of such a bimodal distribution is not still clearly understood, in spite of many previous studies (e.g., Chiappini et al. 1997, 2001; Sch\"onrich \& Binney 2009; Haywood et al. 2016)\footnote[1]{Recently, Grand et al. (2017) based on the latest cosmological hydrodynamical simulation, Auriga, suggested that a gas-rich major merger at the early disk formation phase can lead to a temporal significant star burst event, and as a result a distribution of disk stars becomes naturally bimodal on the $\afe$-$\feh$ plane.}. In our previous work (Toyouchi \& Chiba 2016), we investigated this observational property with the semi-analytic galactic disk evolution model, and suggested based on the model results that to make a bimodal stellar distribution on the $\afe$-$\feh$ plane, a discontinuous radial migration event, as driven by minor merger with a massive satellite, can be an essential mechanism. However, our previous study did not consider the reproduction of the radial dependence of the MDF of the disk stars, and therefore we here reconsider the bimodality on the $\afe$-$\feh$ plane from the current model.

Figure \ref{fig:afe-feh_dist} presents the stellar distributions on the $\afe$-$\feh$ plane produced by our best fit model with color maps and contours. The panels from top left to bottom right represent the stellar distribution observed at the inner ($R$ = 5-7 kpc), solar neighborhood ($R$ = 7-9 kpc), the outer ($R$ = 9-11 kpc), and the summation of the three regions ($R$ = 5-11 kpc), respectively. It follows that our model cannot reproduce a clear bimodal distribution as observed in Hayden et al. (2015), but in $R$ = 5-11 kpc we can find the faint bifurcated distribution around $\feh \sim -0.4$, probably corresponding to the observed high and low-$\afe$ sequences. The high-$\afe$ sequence in our model emerges only at the inner radii, $R$ $<$ 9 kpc, and fades out at the outer radii. Additionally, the metallicity of the density peak on the plane becomes lower along the low-metallicity sequence in the outer disk regions. These properties of the high and low-$\afe$ sequences in our model look similar to those observed for the APOGEE sample, although our models do not provide the dependence of this sequence on the distance from the disk plane that Hayden et al. (2015) actually presented.


\begin{figure}
\begin{center}
\includegraphics[width=10cm,height=12cm]{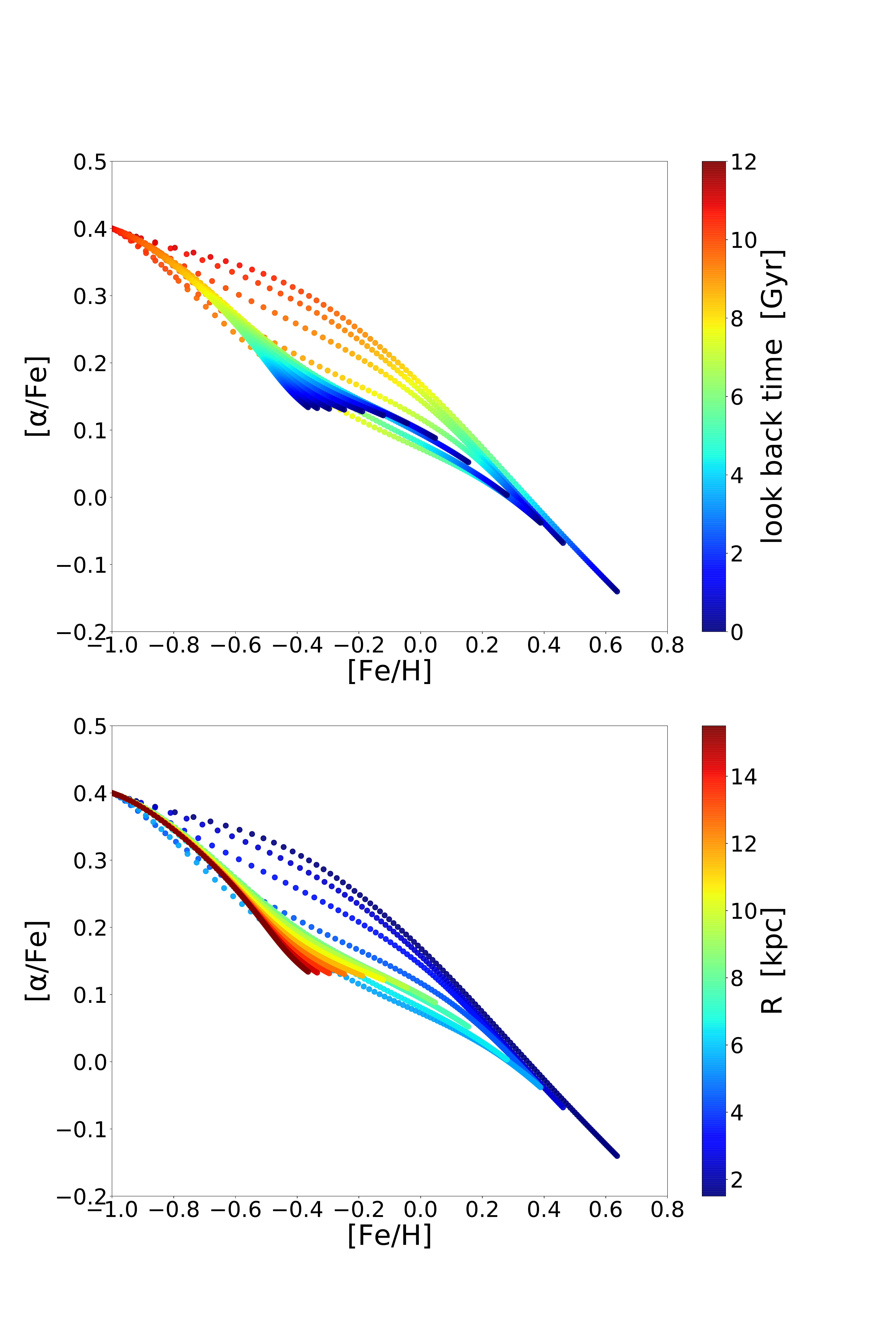}
\end{center}
\caption{$\afe$ and $\feh$ of gas in each radial ring at each time step, respectively. The color of each plot in the top and bottom panels corresponds to the look back time and the radius, respectively.}
\label{fig:afe-feh_plot} 
\end{figure}

To clarify how such a stellar distribution on the $\afe$-$\feh$ plane is formed in our model, in Figure \ref{fig:afe-feh_plot} we plot the $\afe$ vs. $\feh$ for gas in each radial ring at each time step in the model calculation. The color of each plot in the top and bottom panels shows the look back time and the radius, respectively. From the comparison between Figure \ref{fig:afe-feh_dist} and \ref{fig:afe-feh_plot}, we find that the high-$\afe$ and low-$\afe$ sequences obtained in our model generally consist of the old disk stars born in the inner disk regions of $R$ $\lesssim$ 3 kpc at $\gtrsim$ 9 Gyr ago, and the young ones born in the outer disk regions of $\gtrsim$ 6 kpc at $\lesssim$ 7 Gyr ago, respectively. Here, it is worth noting that the decrease of $\afe$ at around $R$ $\sim$ 4-6 kpc is much faster than those in the other regions, thereby the evolutionally paths on the $\afe$-$\feh$ diagram in the inner and outer disk regions are made distinctly different, roughly corresponding to the high and low-$\afe$ sequence. The rapid decrease of $\afe$ is due to the significant increase of the number ratio of SNe Ia to SNe II, $N_\mr{Ia}/N_\mr{II}$, caused by radial migration in which a large amount of intermediate and old disk stars, which eventually explode as SNe Ia, migrate from the inner ($R \lesssim$ 3 kpc) to outer disk regions ($R \gtrsim$ 5 kpc). Thus, the bifurcated distribution on the $\afe$-$\feh$ plane is made in our model.

The suggestion that the high-$\afe$ sequence is originated from the inner disk region and is older than the low-$\afe$ one, has been already made by the previous studies with chemical evolution models, which found that the high and low-$\alpha$ sequences on the $\afe$-$\feh$ plane appear as long as the simulated disk sample is separated by their kinematics or age (e.g., Minchev et al. 2013; Kybryk et al. 2015a). An interesting difference between our and the previous models is that the bimodal distribution can be found for the whole sample in our model, although it looks much weaker than the actually observed one. This difference may be due to the up-bending radial density profile of gas infall, which is newly implied in this study, because our previous model in Toyouchi \& Chiba (2016), which assumed an exponential gas infall profile with no break, cannot show a distinct two sequences for the case of continuous radial migration model (see Figure 9 of the paper). Such an up-bending density profile leads to a more massive inner disk and subsequently more significant transfer of disk stars from inner to outer disk regions than the previous models. Consequently, a significant increase of $N_\mr{Ia}/N_\mr{II}$ realizes in our model, and thereby the clearly distinct high and low-$\afe$ sequences are produced.

Additionally, we note that a similar effect of radial migration on the stellar distribution on the $\afe$-$\feh$ plane is also confirmed in Toyouchi \& Chiba (2016), which can reproduce a bimodal stellar distribution on the $\afe$-$\feh$ plane. According to the previous model calculation, a discontinuous radial migration of disk stars drastically decreases $N_\mr{Ia}/N_\mr{II}$ in the inner disk regions, where the high-$\afe$ sequence stars formed, and temporarily slows down the decrease of $\afe$, consequently leading to the high-$\afe$ peak of stars. Thus adopting not only the stellar distribution for $\feh$, but also for $\afe$ as the observational constraint for model fittings may enable us to reveal the more detailed dynamical evolution history in the Galactic stellar disk, and this refinement is to be explored in future studies.

\subsection{Dependence of radial metallicity gradient of disk stars on their age}

The radial metallicity gradient of the Galactic disk stars has been measured using various tracers including HII regions, star clusters, gaseous nebulae and Cepheid variable stars (e.g., Maciel \& Costa 2009 and references therein). In particular, the field main-sequence turn-off stars are useful tracers, which can provide the time evolution of the radial metallicity gradient (e.g., Nordstr\"om et al. 2004). The most recent observation of the radial metallicity gradients, $\Delta \feh / \Delta R$, of the disk stars as a function of stellar age is provided from the LAMOST survey, which is one of the latest optical spectroscopic survey for the Galactic stars (Xiang et al. 2015). The observation indicates that the radial metallicity gradient is mostly flat for the oldest disk stars with age $>$ 10 Gyr, whereas the younger disk stars with age $<$ 9 Gyr show the negative radial metallicity gradients with the steepest value of $\sim$ $-0.01$ dex/kpc at age $\sim$ 7 Gyr. Such a dependence of radial metallicity gradient of disk stars on their age is generally consistent with the results provided by the previous studies (Casagrande et al. 2011; Toyouchi \& Chiba 2014). 


\begin{figure}
\begin{center}
\includegraphics[width=9cm,height=6cm]{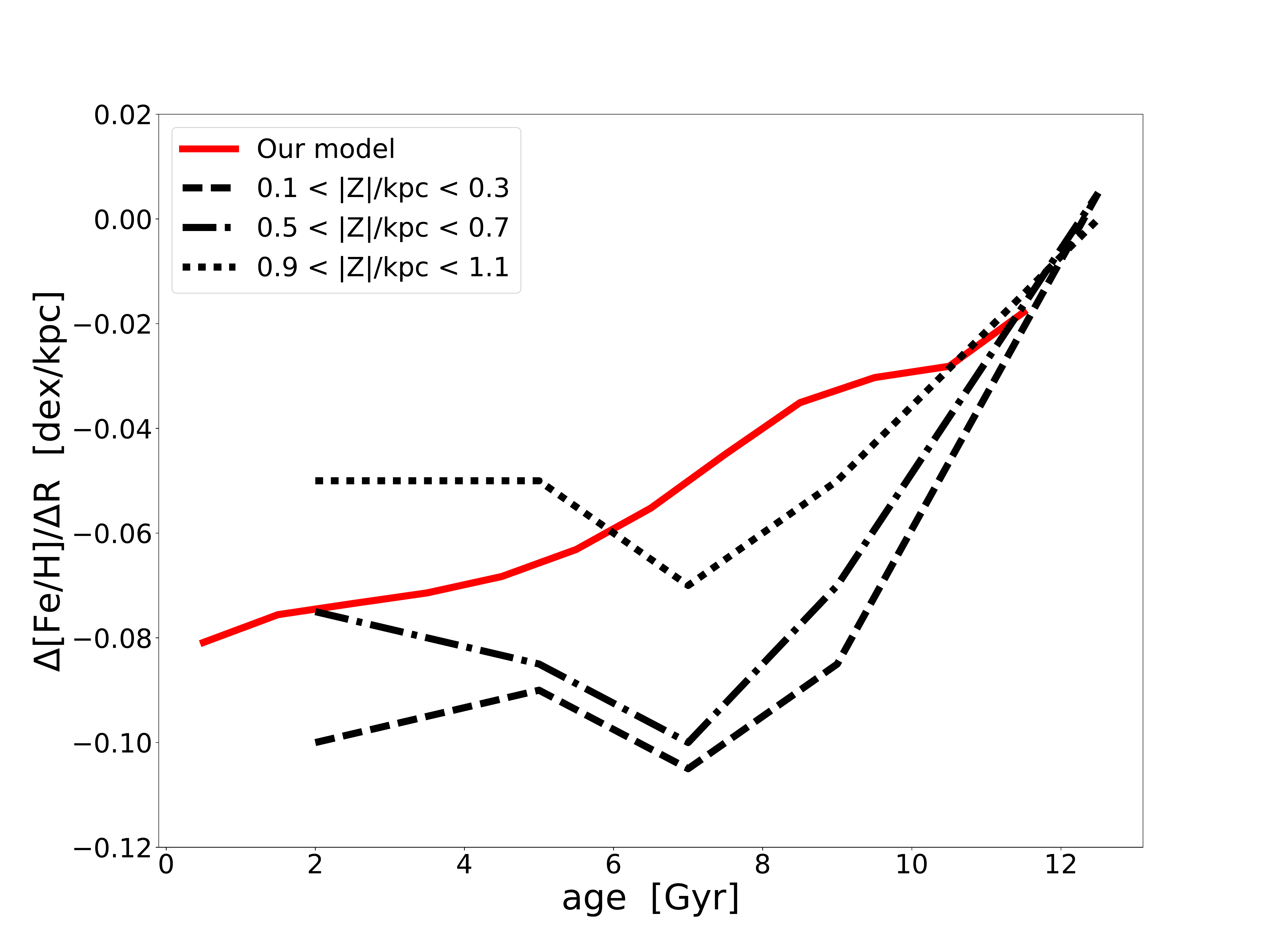}
\end{center}
\caption{The radial metallicity gradient of disk stars as a function of age. The red and three black lines correspond to our model result and the observation for disk stars, locating at the three vertical distances of $|Z|$ = 0.1-0.3, 0.5-0.7, and 0.9-1.1 kpc, presented in Figure 16 of Xiang et al. (2015), respectively.}
\label{fig:rmg-age} 
\end{figure}

We here present the radial metallicity gradient of stars as a function of age obtained in our best fit model in Figure \ref{fig:rmg-age}. The red line shows our model result and for comparison we plot the observed results of Xiang et al. (2015) for the three vertical distances of $|Z|$ = 0.1-0.3, 0.5-0.7, and 0.9-1.1 kpc, because our model result should be regarded as the vertically integrated to the disk plane. We find that the radial metallicity gradient of disk stars in our model gradually steepens with decreasing age from $-0.02$ dex/kpc for the oldest stars to $-0.08$ dex/kpc for the youngest ones. These results appear similar to those obtained in Kubryk et al (2015b). Their chemical evolution model shows that the radial metallicity gradient of stars does not necessarily reflect that of ISM, because the stellar radial migration can wash out the original metallicity gradients, and therefore, the observed metallicity gradient of disk stars are flatter for older disk populations, which have radially well mixed due to their long traveling time. 


\begin{figure*}
\begin{center}
\includegraphics[width=18cm,height=12cm]{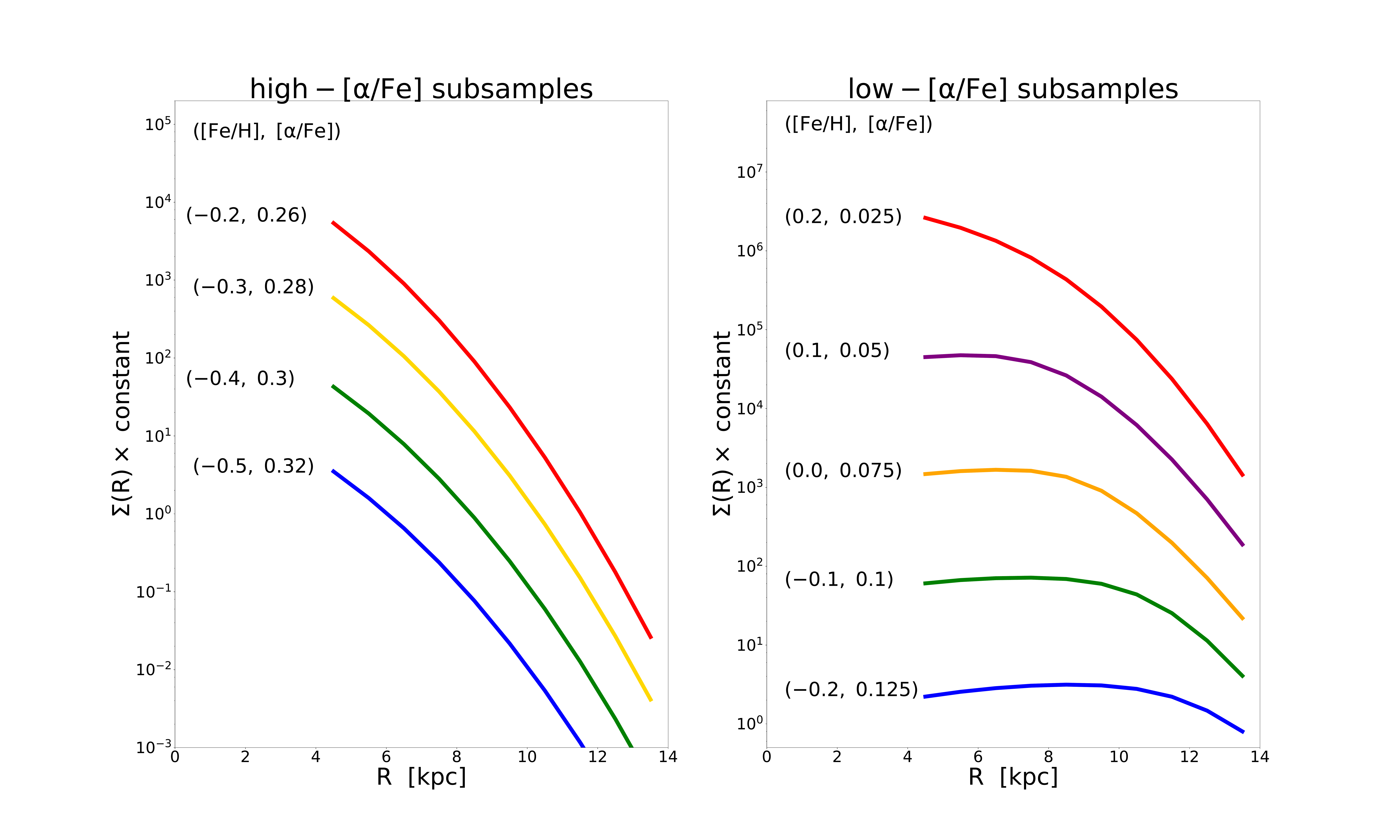}
\end{center}
\caption{The radial density profiles for several abundance-selected subsamples produced by our best fit model. The left and right panels show the radial density profiles for the high and low-$\afe$ subsamples, respectively, and the values of $\afe$ and $\feh$ of each subsample are denoted at the left end of the radial density profile. In this figure, arbitrary offsets in the vertical direction are applied to clearly display the density profile of each subsample. }
\label{fig:density} 
\end{figure*}

We note here that our model can well reproduce the observed metallicity gradient especially for the oldest and youngest populations, but predicts the much flatter radial metallicity gradients at age $\sim$ 5-10 Gyr than those in the observation. This disagreement implies that our chemical evolution model might adopt too simple and perhaps unrealistic assumptions for gas infall, gas outflow, gas re-accretion, and stellar radial migration processes, or still miss some important ingredients in understanding the MW formation. Therefore, we would like to construct more refined models as future studies based on more physically motivated considerations or results of latest numerical simulations. 

\subsection{Radial stellar density profiles as a function of chemical abundance}

The spatial distribution of the Galactic disk stars as a function of stellar age or chemical abundances is a key piece of information to elucidate the formation history of the Milky Way. To obtain this information, Bovy et al. (2016) divided the APOGEE red-clump stars into several subsamples by their $\afe$ and $\feh$, and estimated the spatial density profiles for these abundance-selected subsamples with a careful collection of the observational selection biases in the APOGEE survey. They found based on the analysis that the properties of the density profiles are much different between the high and low-$\afe$ subsamples; the radial profiles for the high-$\afe$ subsamples are well fitted by single exponentials with a common scale length, whereas those for the low-$\afe$ ones are described by broken profiles, for which the radial density gradients are positive and negative at the inner and outer disk regions, and the break radii in the radial profiles move outward with decreasing $\feh$ of subsamples.  Recently, Mackereth et al. (2017) further investigated the disk structure as a function of not only chemical abundance, but also stellar age by using the APOGEE red-clump sample, and suggested that the properties of the high and low-$\afe$ subsamples do not so strongly change with stellar ages. 

The difference of density structures between the high and low-$\afe$ disk populations is expected to reflect the difference of their formation mechanisms. Here, we present the radial density profiles for several abundance-selected subsamples produced by our best fit model in Figure \ref{fig:density}. In this figure, the left and right panels show the radial density profiles for the high and low-$\afe$ subsamples, respectively, and the values of $\afe$ and $\feh$ of each subsample are denoted at the left end of the radial density profile. We find from this figure that our model can remarkably reproduce the general properties of the observed radial density profiles for both the high and low-$\afe$ subsamples. This result is not so surprising because our model can also reproduce the stellar distribution on the $\afe$-$\feh$ plane and its spatial dependence. As noted in Section 5.1, the high and low-$\afe$ disk stars formed at $R$ $<$ 3 kpc and $>$ 6 kpc, respectively, and they have radially migrated inward and outward from their birth radii. Therefore, the present stellar density profiles of the high-$\afe$ subsamples monotonically decreases with increasing radii at $R$ $>$ 4 kpc, whereas those of the low-$\afe$ ones have the peaks around at the radii, roughly corresponding to the average birth radii for the member stars of the subsample, thereby the peak radii becomes larger for more metal-poor subsamples.

Thus, our chemical evolution model enables us to understand not only the radial dependence of the MDF of disk stars, but also the rough trends of the other several important observations of the Galactic stellar disk. Therefore, we consider that our chemical evolution model, which is consistent with the observable constraints, well captures important aspects of the complex formation history of the Milky Way.

\section{SUMMARY \& CONCLUSION} \label{sec:summary}

In this paper, to get new insights into the formation history of the stellar disk in the Milky Way, we have attempted to reproduce the radial dependence of the metallicity distribution function of the Galactic disk stars with the semi-analytic chemical evolution model. Our model includes the effects of gas infall, re-accretion of gas once ejected from the disk, and radial migration processes in galaxy evolution, each of which is characterized by model parameters. We have determined the best set of model parameters, which fit to the present structural and chemical properties of the Milky Way stellar disk based on the Markov Chain Monte Carlo method. We have succeeded to find the solution to this galaxy formation model, which reproduces the observed radial dependences of the mean, standard deviation, skewness, and kurtosis of the metallicity distribution functions of disk stars. We have also derived various fundamental results, and obtained the following new implications for gas infall, re-accretion of outflowing gas, and radial migration processes in the formation history of the Milky Way.
\begin{itemize}

\item The time scale of gas infall is shorter in the inner disk regions, which is associated with a shorter gas cooling and collapsing time in the inner halo. The total surface mass density of gas infall follows an up-bending profile with the break radius of $\sim$ 5 kpc, and such a profile is expected to reflect the angular momentum distribution of infalling gas on the galactic disk plane. As a result, gas accretion onto the disk regions below $R \sim$ 5 kpc occurs very rapidly at the early disk formation phase, whereas gas accretion onto outer radii proceeds slowly over the whole disk formation phase. This gas infall history leads to the inside-out galactic disk evolution.

\item Outflowing gas ejected from the Galactic disk falls back preferentially onto the outer disk parts, and such a gas re-accretion process is an essential contributor of gas supply into those regions of the Galactic disk. We propose that this process provides important effects on the structural and chemical evolution of the MW. As a result of the combination of the re-accretion of metal-enriched gas with the inside-out galactic disk evolution, the inner disk regions of the MW formed very rapidly from metal-poor gas, whereas the outer regions have been gradually constructed from metal-rich gas supplied from re-accretion of once ejected gas, consequently forming the observed narrower metallicity distribution functions of disk stars at larger radii.

\item The net radial transport of metal-rich disk stars from the inner to outer disk regions produces significant high-metallicity tails in the metallicity distribution functions in the outer disk regions, as observed in the Galactic stellar disk. Therefore, the radial migration process is essential to reproduce the radial dependence of the skewness of the metallicity distribution functions of the disk stars.

\end{itemize}

Moreover, based on the further analysis of our best fit model, we have found that our model reasonably reproduces other various important observational properties of the Galactic stellar disk, e.g., the $\afe-\feh$ relation and the stellar density profiles as a function of $\afe$ and $\feh$. Therefore, our chemical evolution model can be applied to the description of galaxy evolution. In future, we will improve our model to include more realistic descriptions of, not only the processes taken into account in this study, but also stellar dynamics in order to compare with the detail kinematic information given by the forthcoming data release of the Gaia survey. Additionally, we will further apply this model to other relevant subjects, including the detail analysis of the SEDs of local star-forming disk galaxies and the subject of the galactic habitable regions in the Milky Way. Thus, our chemical evolution model would play as an invaluable tool in studying various subjects relating to the evolutions of disk galaxies.

\acknowledgments
We are grateful to the referee for constructive comments which have helped us improve our paper substantially. This work is supported in part by JSPS Grant-in-Aid for Scientific Research (No. 27-2450 for DT) and MEXT Grant-in-Aid for Scientific Research (No. 15H05889, 16H01086 and 17H01101 for MC).

\clearpage

\end{document}